\documentclass[review]{elsarticle}

\usepackage{lineno,hyperref}
\usepackage[T1]{fontenc}
\modulolinenumbers[5]

\journal{Journal of \LaTeX\ Templates}

\begin{document}

\begin{frontmatter}

\title{Small Bodies Near and Far (SBNAF): a benchmark study on physical and thermal
properties of small bodies in the Solar System\tnoteref{mytitlenote}}
\tnotetext[mytitlenote]{The research leading to
these results has received funding from the European Union's Horizon 2020 Research
and Innovation Programme, under Grant Agreement no 687378.}

\author[add1]{T.\ G.\ M\"uller\corref{correspondingauthor}}
\cortext[correspondingauthor]{Corresponding author}
\ead{tmueller@mpe.mpg.de}
\author[add2]{A.\ Marciniak}
\author[add3]{Cs.\ Kiss}
\author[add4]{R.\ Duffard}
\author[add1]{V.\ Al\'{i}-Lagoa}
\author[add2]{P.\ Bartczak}
\author[add2]{M.\ Butkiewicz-B\k{a}k}
\author[add2]{G.\ Dudzi{\'n}ski}
\author[add4]{E.\ Fern{\'a}ndez-Valenzuela}
\author[add3]{G.\ Marton}
\author[add4]{N.\ Morales}
\author[add4]{J.-L.\ Ortiz}
\author[add2]{D.\ Oszkiewicz}
\author[add2]{T.\ Santana-Ros}
\author[add3]{R.\ Szak{\'a}ts}
\author[add4]{P.\ Santos-Sanz}
\author[add3]{A.\ Tak{\'a}csn{\'e} Farkas}
\author[add3]{E.\ Varga-Vereb{\'e}lyi}

\address[add1]{Max-Planck-Institut f\"ur extraterrestrische Physik (MPE), Giessenbachstrasse 1, 85748 Garching, Germany.}
\address[add2]{Astronomical Observatory Institute, Faculty of Physics, A.\ Mickiewicz University, S{\l}oneczna 36, 60-286 Pozna{\'n}, Poland.}
\address[add3]{Konkoly Observatory, Research Centre for Astronomy and Earth Sciences, Hungarian Academy of Sciences, H-1121 Budapest, Konkoly Thege Mikl{\'o}s {\'u}t 15-17, Hungary.}
\address[add4]{Instituto de Astrof{\'i}sica de Andaluc{\'i}a (CSIC), Glorieta de la Astronom{\'i}a s/n, 18008 Granada, Spain.}




\begin{abstract}
The combination of visible and thermal data from the ground and
astrophysics space missions is key to improving the
scientific understanding of near-Earth, main-belt, trojans, centaurs, and trans-Neptunian objects.
To get full information on a small sample of selected bodies we combine different
methods and techniques: lightcurve inversion, stellar occultations, thermophysical
modeling, radiometric methods, radar ranging and adaptive optics imaging.
The SBNAF project will derive size, spin and shape, thermal inertia,
surface roughness, and in some cases bulk densities and even internal
structure and composition, for objects out to the most distant regions in
the Solar System. The applications to objects with ground-truth information
allows us to advance the techniques beyond the current
state-of-the-art and to assess the limitations of each method.
We present results from our project's first phase: the analysis of
combined Herschel-KeplerK2 data and Herschel-occultation data for TNOs;
synergy studies on large MBAs from combined high-quality visual and
thermal data; establishment of well-known asteroids as celestial 
calibrators for far-infrared, sub-millimetre, and
millimetre projects; first results on near-Earth asteroids
properties from combined lightcurve, radar and thermal measurements,
as well as the Hayabusa-2 mission target characterisation.
We also introduce public web-services and tools for studies of small bodies
in general.
%
\end{abstract}


\end{frontmatter}


\section{Introduction}
\begin{figure}[h!tb]
 \resizebox{\hsize}{!}{\includegraphics{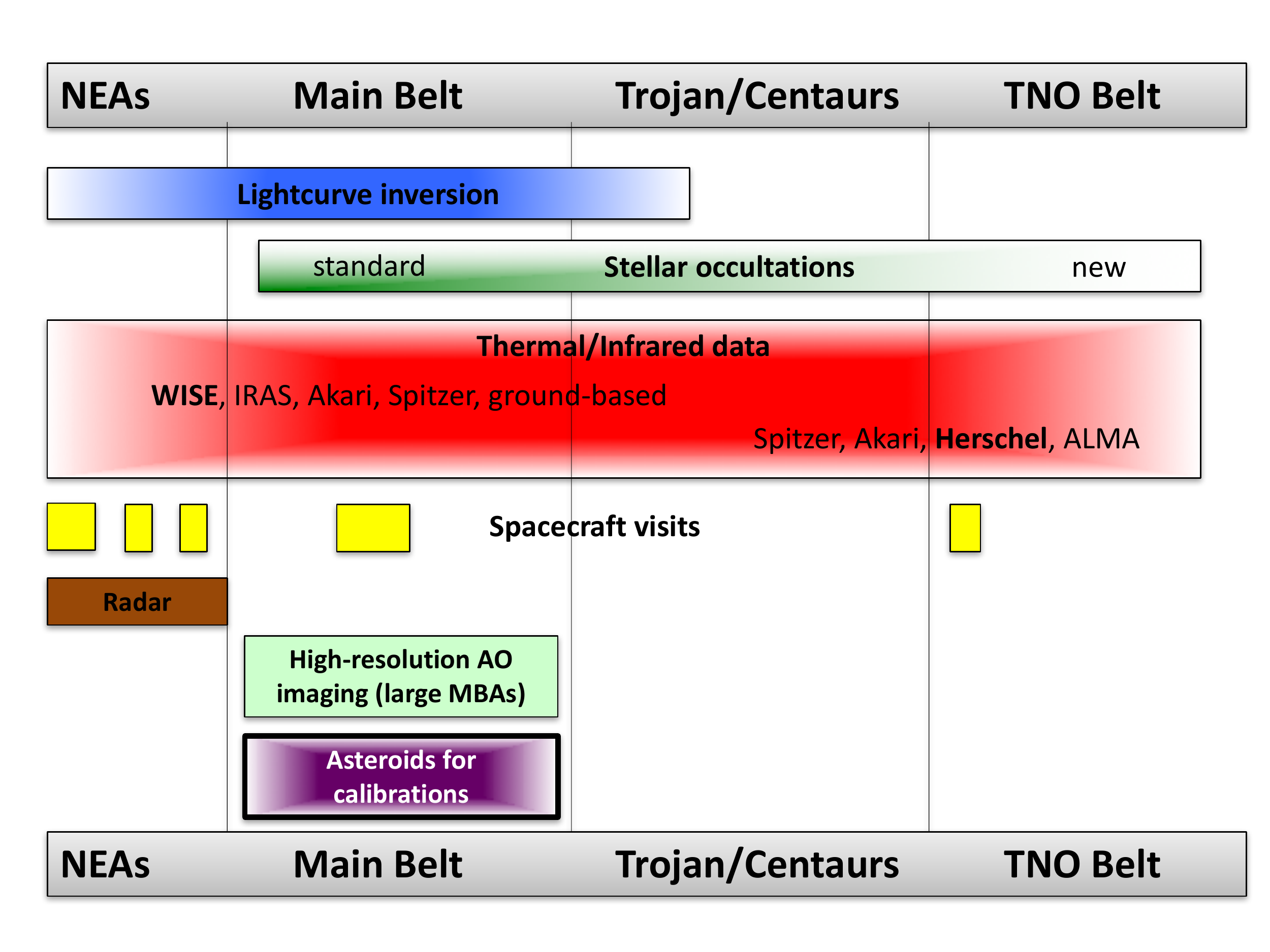}}
  \caption{Overview of the different techniques applied to minor
           bodies at different distances from the Sun. The range where
           a given technique can be used is very restricted,
           making the reconstruction of object properties more complex
           and strongly dependent on the availability of suitable data.
     \label{fig:alltechniques}}
\end{figure}

We conduct an EU Horizon 2020\footnote{Research and Innovation programme of
the European Union, see {\tt http://www.horizont2020.de/}}-funded
benchmark study\footnote{Scientific exploitation of astrophysics, comets, and planetary data:\\
{\tt https://ec.europa.eu/\-research/\-participants/\-portal/\-desktop/\-en/\-opportunities/h2020/\-topics/\-compet-05-2015.html}}
(2016-2019) that addresses critical points in reconstructing physical and thermal
properties of near-Earth (NEA), main-belt (MBA), and trans-Neptunian objects (TNO).
The combination of the visual and thermal data from ground and
astrophysics space missions like Hubble\footnote{The Hubble Space Telescope, see {\tt http://hubblesite.org/}},
Kepler-K2\footnote{The Kepler Space Telescope, see {\tt https://keplerscience.arc.nasa.gov/}},
WISE/NEOWISE\footnote{Wide-field Infrared Survey Explorer, see {\tt https://neowise.ipac.caltech.edu/}},
IRAS\footnote{Infrared Astronomical Satellite, see {\tt http://irsa.ipac.caltech.edu/IRASdocs/iras.html}},
Herschel\footnote{The Herschel Space Observatory, see {\tt http://sci.esa.int/herschel/}},
Spitzer\footnote{The Spitzer Space Telescope, see {\tt http://www.spitzer.caltech.edu/}},
AKARI\footnote{The Infrared Imaging Satellite "AKARI" (ASTRO-F), see {\tt http://global.jaxa.jp/\-projects/\-sat/\-astro\_f/}},
and others, is key to improving the scientific understanding of these objects.
The development of new and improved tools is crucial for the interpretation
of much larger data sets from WISE/NEOWISE,
Gaia\footnote{Astrometric space observatory of the European Space Agency, see {\tt http://sci.esa.int/\-gaia/}},
JWST\footnote{James Webb Space Telescope, see {\tt https://www.jwst.nasa.gov/}},
or NEOShield-2\footnote{EU-funded project on "Science and Technology for Near-Earth
Object Impact Prevention, see {\tt http://www.neoshield.eu/\-science-technology-asteroid-impact/}}.
Some of our results will be used in support of the operation of
interplanetary missions, and for the exploitation of in-situ data.
Depending on the availability of data, we combine different methods
and techniques to get full information on selected bodies.
Figure~\ref{fig:alltechniques} shows the typical data and applicability
of techniques as a function of distance from the Sun:
lightcurve inversion, stellar occultations, thermal/infrared data (for
thermophysical modeling and radiometric methods), radar ranging and adaptive optics imaging.
The applications to objects with ground-truth information
from interplanetary missions Hayabusa\footnote{Asteroid Explorer of the Japan Aerospace
Exploration Agency (JAXA), see {\tt http://global.jaxa.jp/\-projects/\-sat/\-muses\_c/}},
NEAR-Shoemaker\footnote{The Near Earth Asteroid Rendezvous – Shoemaker mission,
see {\tt https://solarsystem.nasa.gov/\-missions/\-near}},
Rosetta\footnote{Comet mission by ESA, see {\tt http://sci.esa.int/\-rosetta/}},
DAWN\footnote{Space probe to the asteroids Vesta and Ceres, see {\tt https://dawn.jpl.nasa.gov/}},
or New Horizons\footnote{NASA mission to Pluto, see {\tt http://pluto.jhuapl.edu/}}
(see yellow blocks in Fig.~\ref{fig:alltechniques})
allows us to advance the techniques beyond the current state-of-the-art
and to assess the limitations of each method.
Another important aim is to build accurate thermophysical asteroid models
to establish new primary and secondary celestial calibrators for the far-infrared (far-IR),
sub-millimeter (submm), and millimeter (mm) range
(ALMA\footnote{The Atacama Large Millimeter/submillimeter Array, see {\tt http://www.almaobservatory.org/}},
SOFIA\footnote{Stratospheric Observatory for Infrared Astronomy, see {\tt https://www.sofia.usra.edu/}},
APEX\footnote{Atacama Pathfinder EXperiment, see {\tt http://www.apex-telescope.org/}},
IRAM\footnote{Radio telescopes operated by the Institute for Radio Astronomy in the
Millimeter Range (IRAM), see {\tt http://www.iram-institute.org/}},
and others), as well as to provide a link to the high-quality
calibration standards of Herschel and
Planck\footnote{ESA's Planck space telescope, see {\tt http://sci.esa.int/\-planck/}}.
The SBNAF project will derive
physical, thermal, and compositional properties for small bodies 
throughout the Solar System. The target list comprises
recent interplanetary mission targets, two samples of main-belt
objects, representatives of the Trojan and Centaur populations, and all
known dwarf planets (and candidates) beyond Neptune.

We introduce the relevant observing techniques (Sect.~\ref{sec:techniques}),
our target sample (Sect.~\ref{sec:targets}) and the related science
questions. Selected results from our project's first year will be
discussed in Section~\ref{sec:synergies}, new tools and web-services
for studies of small bodies will be presented in Section~\ref{sec:results}.
We conclude with an outlook (Sect.~\ref{sec:outlook}) for the next project phases.

\section{Observing techniques}
\label{sec:techniques}

Small bodies are typically not resolved and models are needed to translate
disc-integrated signals into physical properties. The SBNAF project will take 
advantage of existing observational data taken remotely from ground-based
observatories and astrophysical missions, but we will also conduct (mainly
photometric) measurements of a set of selected targets to vitally enhance
the amount of information we can derive from them. 


\subsection{Lightcurves in the visible}
Lightcurve inversion techniques are used to find an object's rotation period,
its shape and spin-axis orientation. It requires the availability of
multi-epoch and multi-apparition lightcurve measurements of sufficient quality.
These kind of data are only available for NEAs, MBAs, and very few more 
distant bodies (see Fig.~\ref{fig:alltechniques}).

\begin{figure}[h!tb]
 \resizebox{6cm}{!}{\includegraphics{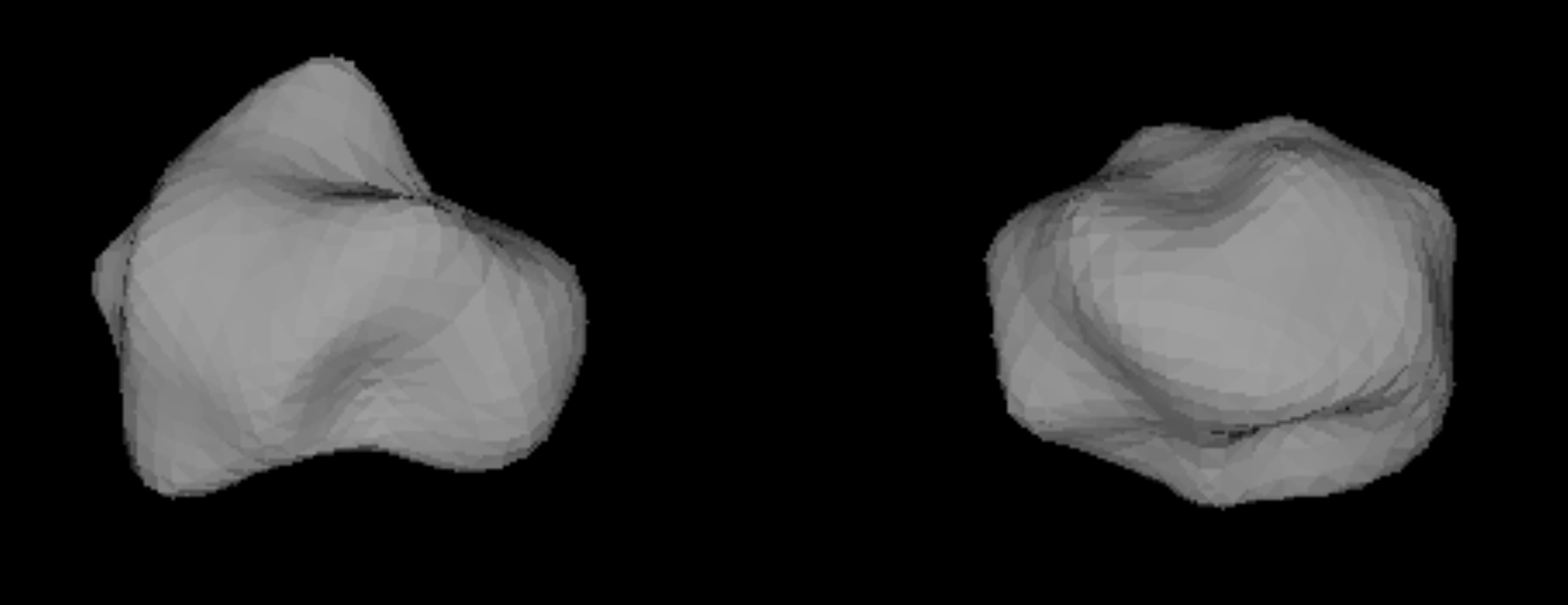}}
 \vspace*{-3cm}

 \resizebox{6cm}{!}{\includegraphics{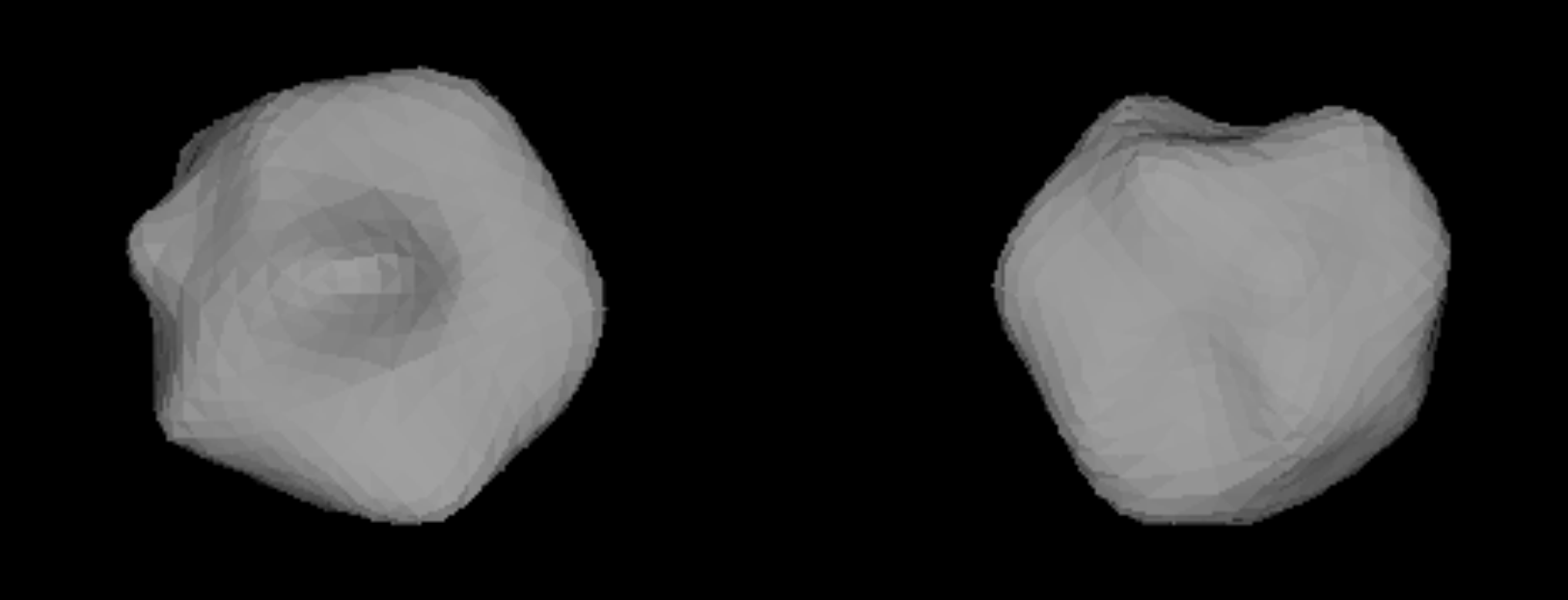}}
 \resizebox{6cm}{!}{\includegraphics{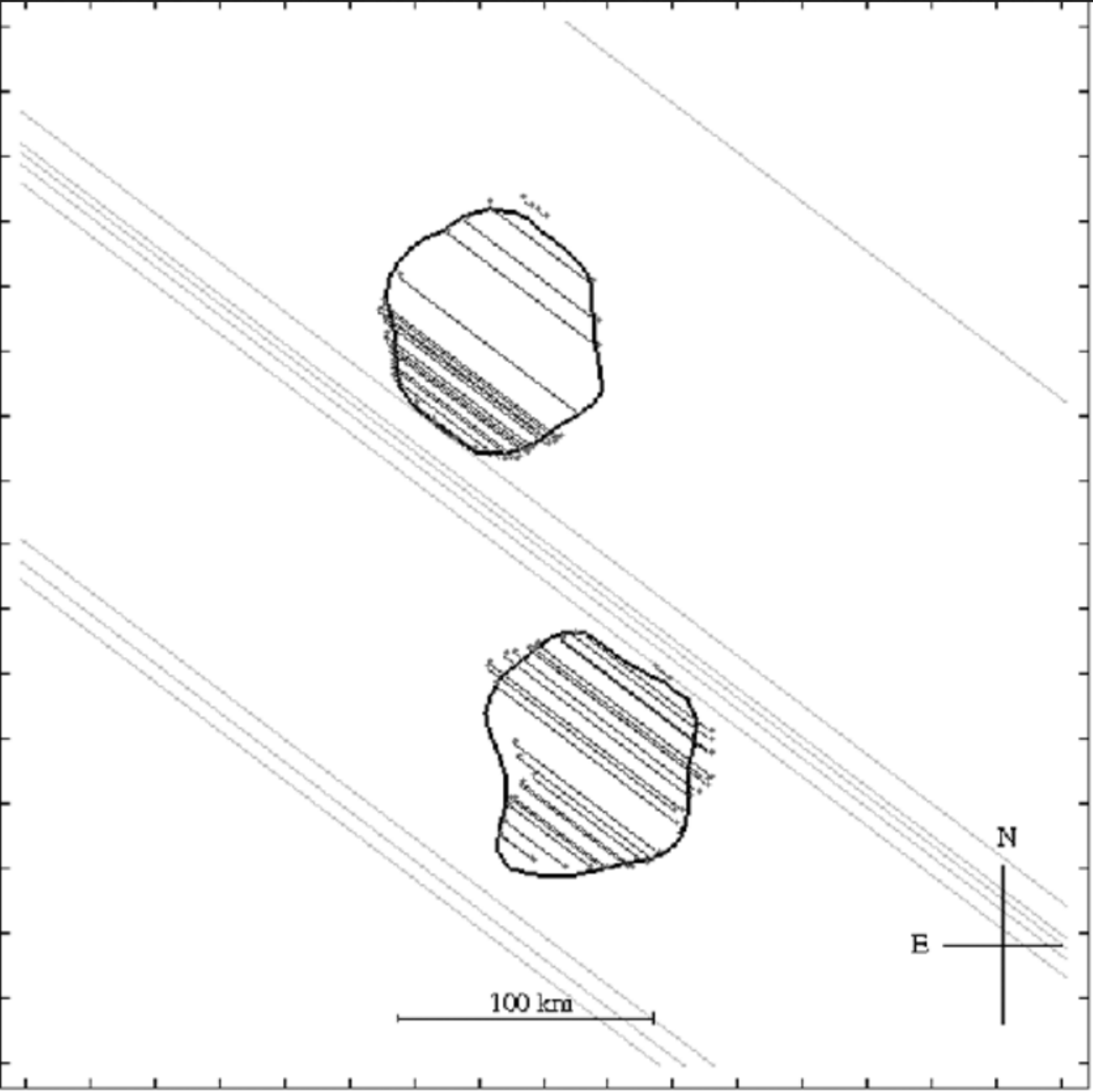}}

  \caption{Left: Shape models for two components of the binary asteroid (90) Antiope 
           obtained with the inversion of lightcurves using the SAGE algorithm 
           for non-convex shapes (Bartczak et al.\ \cite{bartczak14}) in an equatorial view
           (top) and polar view (bottom).
           Right: A comparison between the stellar occultation chords and the projected
           non-convex shape solution.
     \label{fig:lc_shapes}}
\end{figure}

Traditional, dense lightcurves are today available for over 10\,000 asteroids.
The data sets are stored in the LCDB\footnote{Asteroid Light Curve Database at
{\tt http://alcdef.org/}} database (Warner et al.\ \cite{warner09})
which is being regularly updated. Apart from continuous lightcurves, also an
increasing number of so called "sparse data" recently appear, which are sparse
in time absolute photometric measurements, usually obtained as a result or
byproduct of wide-field surveys. In some cases the latter allow for finding
rotational periods (Waszczak et al.\ \cite{waszczak15}), or spin axis position
determinations (\v{D}urech et al.\ \cite{durech16}) for a large number of tagets.

Sparse lightcurve data are scientifically interesting (e.g., for searching for
close and contact binaries via their diagnostic large lightcurve amplitude;
Sonnett et al.\ \cite{sonnett15}), but the shape information content is very limited.
For precise shape reconstruction there are much more stringent demands for
the lightcurve data, which should be dense, low-noise, and come from a wide
range of viewing geometries. Consequently, in order to reconstruct detailed
shapes of asteroids, well coordinated campaigns are needed. The key is to
complement already available data (stored in e.g.\  ALCDEF\footnote{The Asteroid
Lightcurve Data Exchange Format, see {\tt http://alcdef.org/}},
or in APC\footnote{The Asteroid Photometric Catalog, see
{\tt http://asteroid.astro.helsinki.fi/\-apc}} databases) with
observations in different geometries, to probe the lightcurve changes over
various aspect and phase angles. Because of the high demand of observing time,
precise shape models can only be obtained for a small number of asteroids.

The field of spin and shape modelling of asteroids has seen a huge development in
recent years. Since the introduction of the lightcurve inversion technique
at the beginning of the last decade (Kaasalainen \& Torppa \cite{kaasalainen01a};
Kaasalainen at al.\ \cite{kaasalainen01b}), over 900 spin and shape models have been
published (e.g., Hanu\v{s} at al.\ \cite{hanus13}; Marciniak et al.\ \cite{marciniak12}),
usually based on sparse data. First attempts of multi-data inversion have been made
in the last years (e.g., KOALA code, Carry et al.\ \cite{carry12};
ADAM algorithm, Viikinkoski et al.\ \cite{viikinkoski15}).
Previously obtained shape models have also been size-scaled using data from
stellar occultations (\v{D}urech et al.\ \cite{durech11}). In this way, it was shown
that many inversion solutions fit the data from independent methods. It has also
been demonstrated recently, that lightcurves alone contain enough information for
reliable non-convex modelling (the SAGE algorithm, Bartczak et al.\ \cite{bartczak14,bartczak17};
see Fig.~\ref{fig:lc_shapes}).

Today, there is a possibility to join many types of complementary data to
construct full physical models of asteroids, which would be an invaluable
cornerstone for calibration of various methods and extrapolating gained knowledge to
the whole range of objects, especially those with less rich available data sets.
For example with thermal infrared (IR) data we see emission influenced by the thermal inertia
and roughness of the surface and also sub-surface emission at submm/mm wavelengths,
which makes a direct comparison with optical lightcurves more complex (see also
discussion in M\"uller et al.\ \cite{mueller17b}). Thermal data may also bear
contributions from non-illuminated, yet warm parts of the surface (e.g., Nuggent
et al.\ \cite{nugent17}). Studying the relation of these two types of data on the
basis of a few well-studied objects will help in developing a tool with great
potential to infer information on albedo, size, spin, thermal inertia, and
large-scale surface and regolith characteristics.

Physical properties of asteroid surfaces are the missing link in e.g.\ YORP effect modelling,
which has been shown to change the spin frequencies and spin axis positions of small
and medium-sized asteroids (Vokrouhlicky et al.\ \cite{vokrouhlicky03}).
However, widely applicable
small-body modelling
techniques based on such
varied sources of data (optical and thermal) is still missing. Thus we are
going to address these issues in the present project. This way we will establish
strong foundations for further studies of asteroid physical properties.

\subsection{Radar technique}
Radar is a technique used to retrieve information about asteroids.
Its uniqueness lies in the observer's control of the transmitted signal, which is not the case in other
techniques, like photometry. Thus, radar observations can be described as an experiment
(Ostro et al.\ \cite{ostro02}, and references therein). 

Only NEAs and MBAs that pass sufficiently close to Earth can be observed by radar, as the
signal from the telescope fades very quickly with distance (see Fig.~\ref{fig:alltechniques}).
Signals are sent from radio telescope and they bounce off the surface
of the target body to be then recorded back on Earth. Asteroids come in variety of shapes,
and the signal travels different distance depending on the part of asteroid that it hits.
The echo (returning signal) arrives at different times, which can be translated into
a size estimate.
Since asteroids rotate, we can use this phenomena and take advantage of the Doppler effect.
When the light emitting (or reflecting) objects (or surface elements) move towards the observer
the registered light will have higher frequency ("blue-shifted") than the emitted (reflected) light.
Similarly, if the object is moving away, its frequency is lower, and the light
is "red-shifted". When an asteroid rotates, surface elements farther from the spin axis move at 
higher velocities.
As a result, every pixel on radar-echo image is an intensity of returning signal at given time delay
(distance) and frequency (velocity).

If we want to use radar images to model an asteroid, we have to simulate radar observations
of a model object and compare it with observations. The model object is then modified
in complicated patterns until the model echo matches the observed echo. This method works
best when the object's spin properties are known and sufficient good-quality visible 
lightcurves are available.

\subsection{Occultations}
\begin{figure}[h!tb]
 \resizebox{6cm}{!}{\includegraphics{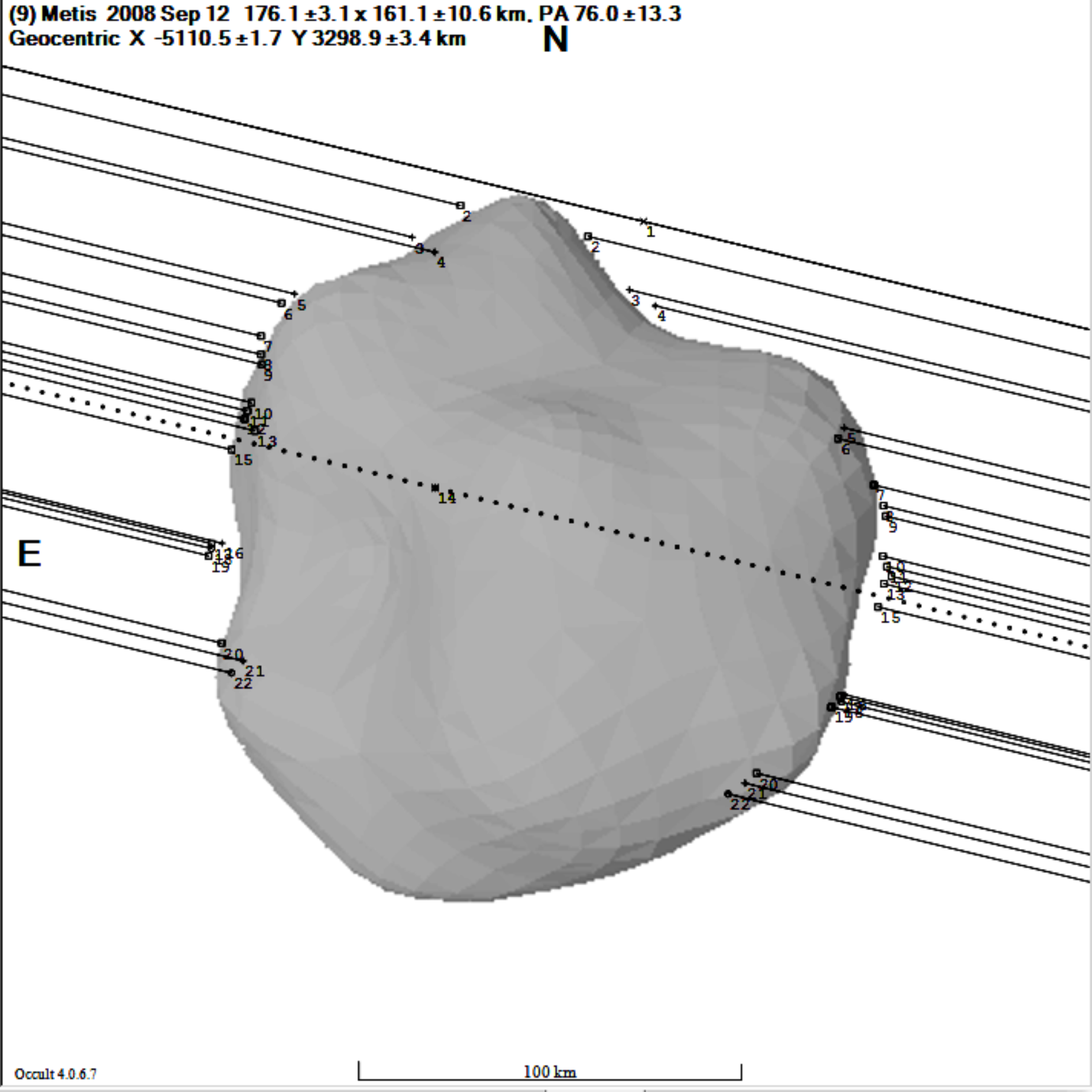}}
 \resizebox{6cm}{!}{\includegraphics{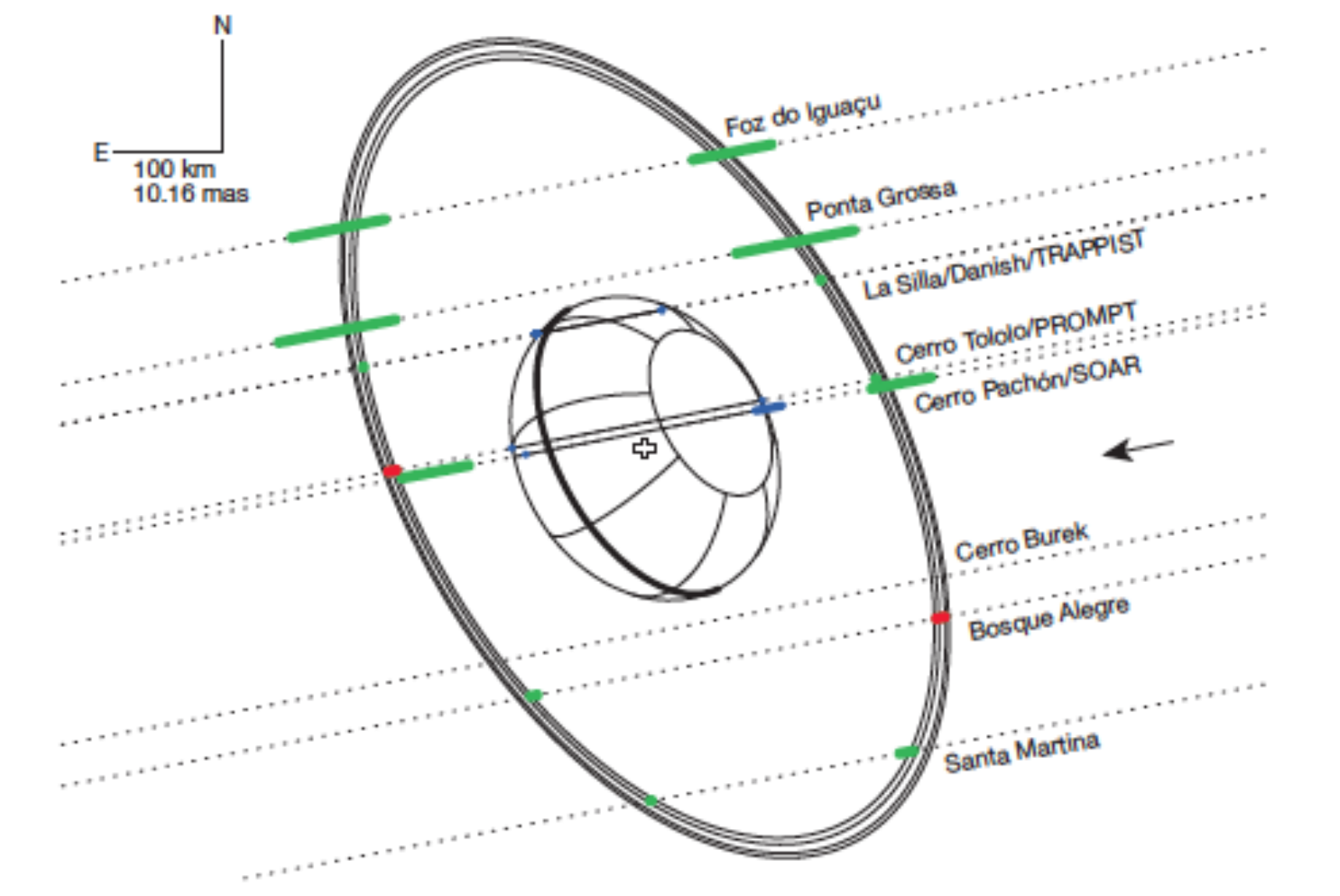}}
  \caption{Left panel: dense telescopic observations (chords) of the stellar
          occultation by (9) Metis revealing the shape of the occulting body,
          with the superimposed independent shape model based exclusively on
          lightcurves inversion technique (Bartczak et al.\ \cite{bartczak14a}).
          Right panel: observations of the stellar occultation by (10199)
          Chariklo, where the presence of rings around a minor body was
          discovered for the first time (Braga-Ribas et al.\ \cite{braga-ribas14}).
     \label{fig:occultation}}
\end{figure}

It is a simple measurement technique to derive the size and the projected cross section of a small body
in a direct way. The basis is to predict when the particular body will pass in front of a certain star. 
One simply measures the flux of the star before, during and after the occultation
from a few locations on Earth within the predicted shadow.
It provides area-equivalent diameters with kilometric accuracy and
it allows to determine the projected shape (a 2-D snapshot) of the body. It can reveal the presence
of atmospheres with pressures down to a nano-bar (nbar) level, discover possible satellites, rings
or material orbiting around a given object (see Fig.~\ref{fig:occultation}).
Stellar occultations of planets, satellites (including the Moon) and
also minor bodies (Elliot \cite{elliot79}; Elliot \& Young \cite{elliot92}) have been
recorded over the last decades.
This technique is well developed for these bodies, but it is only an emerging field
for TNOs and Centaurs. Predicting and observing stellar
occultations by TNOs is extremely difficult and challenging because the angular diameters of
TNOs are very small and neither the stellar catalogues nor the TNOs orbits have the
accuracy required to make reliable predictions well in advance.

A multi-chord stellar occultation by TNOs allows us to determine the
projected shape and orientation of the body in the plane of the sky at the moment of the
occultation. However, this information is insufficient to determine the true 3D shape
of the body and its spin axis orientation in space. Combining the occultation-derived
information with rotational lightcurves one can distinguish whether the 3D shape of
the body is an oblate spheroid or a Jacobi ellipsoid. Usually, a very low TNO lightcurve
amplitude implies a MacLaurin oblate shape, whereas amplitudes larger than 0.15 mag
imply Jacobi shapes (Duffard et al.\ \cite{duffard09}). But the spin axis orientation is still not well constrained
(unless many high-precision rotational lightcurves spanning many years exist).
Modelling thermal observations can be of great help in this regard. It
allows us to put tighter constraints on the spin axis orientation
by modelling the thermal output of the object. The basic parameters that can be
constrained with thermophysical models are the size, shape, albedo, rotation rate
(sometimes even the direction of rotation), the spin-axis orientation,
and surface properties such as e.g.\ thermal inertia. Given that the occultation
timings provides a very accurate size, shape and albedo, and if the rotation
period is also known from the rotational lightcurves, the remaining parameters
can be tightly constrained by combining these techniques. Thus, the combination
of occultation results, optical
lightcurves and thermal measurements allows us to reconstruct a full 3D shape and
spin axis orientation in space. Once this all is known, bulk densities can be
determined accurately using the Chandrasekhar figures of equilibrium formalism.
This works very well for icy dwarf planets (mostly large TNOs) which are expected
to be in hydrostatic equilibrium.

One example of the power of this technique is brought by the work of Ortiz et al.\
(\cite{ortiz12}), who found a radius of the TNO named Makemake to be 1430 $\pm$ 9\,km,
and a hint of an atmosphere. Recently, the existence of rings around two Centaurs
has been discovered by means of stellar occultation: around (10199) Chariklo by
Braga-Ribas et al.\ (\cite{braga-ribas14})
and around (2060) Chiron by Ortiz et al.\ (\cite{ortiz15}).

\subsection{Radiometric technique}
This technique refers to the determination of the radius of the small body
by fitting thermal emission models to observed thermal flux densities. The
first applications of these techniques date back to the 1970's (for a recent
review, see Delbo et al.\ \cite{delbo15}). 
In a nutshell, the warmer a body is, the higher its emitted flux needs to be
in order to stay in thermal equilibrium.
Main belt asteroid surfaces are around 300\,K and
are best observed at around 10\,$\mu$m (data from ground and space),
TNOs surfaces are at around 30 - 40\,K so the best wavelength
range to observe them is between 70 to 100\,$\mu$m (data are mainly coming from space projects). 
Two main radiometric methods allow the exploitation of mid- and far-infrared thermal data with the goal
to obtain size and albedo of asteroids: the Standard Thermal Model (STM;
Lebofsky et al.\ \cite{lebofsky86}), and the near-Earth asteroid thermal model (NEATM; Harris \cite{harris98}).
On the other hand, if the shape and rotational properties of the object are
known, we can model instantaneous surface temperatures accounting for the
heat conductivity of the material as well as surface roughness. These are
typically called ``thermophysical models'' (TPM; see references in 
Harris \& Lagerros \cite{harris02} or Delbo et al.\ \cite{delbo15}).
If the shapes have no
absolute scale, as it is the case for those derived from light curve inversion
techniques for example, TPMs can help find the best scaling factors.
The corresponding volume can be used to find more reliable equivalent
diameters, or densities in cases where the asteroid mass is known. 
If multi-epoch thermal data are available for a given object, then it is possible to
derive reliable thermal properties (thermal inertia, thermal conductivity),
to estimate grain sizes on the surface or to do a simple study 
on the expected surface roughness (see Delbo et al.\ \cite{delbo15} and references therein).
The radiometric techniques work for all IR-detectable bodies in the Solar System
(see Fig.~\ref{fig:alltechniques}). A recent example for radiometric applications
for a large sample of Mars-crossing asteroids was presented by Al{\'i}-Lagoa et al.\
(\cite{alilagoa17}). The "TNOs-are-cool" Herschel Space Observatory
Key project (a large Herschel project with more tan 370\,h of granted time) has
gathered thermal data for more than 130 TNOs (M\"uller et al.\
\cite{mueller09}; Kiss et al.\ \cite{kiss13}; Lellouch et al.\ \cite{lellouch13}; Lacerda et al.\ \cite{lacerda14}).
A good example of capabilities of the radiometric method
based on Herschel observations is a study of the very distant (88\,AU) TNO named Sedna.
P\'al et al.\ (\cite{pal12}) derived a diameter of 995 $\pm$ 80\,km and geometric
albedo of 0.32 $\pm$ 0.06. Sedna is not easily accessible otherwise.

\begin{figure}[h!tb]
 \resizebox{6cm}{!}{\includegraphics{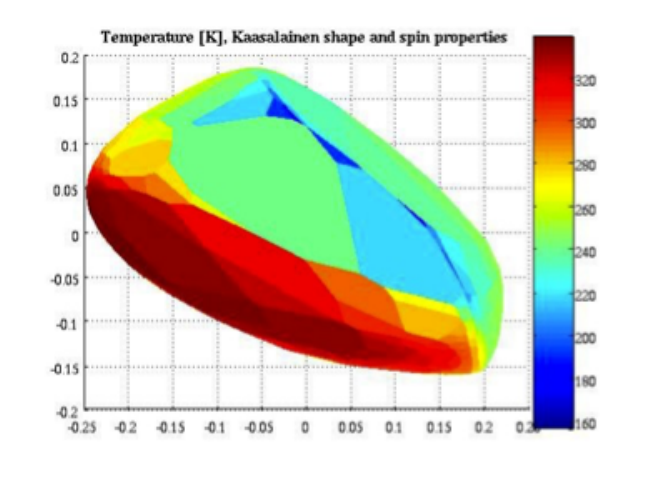}}
 \resizebox{6cm}{!}{\includegraphics{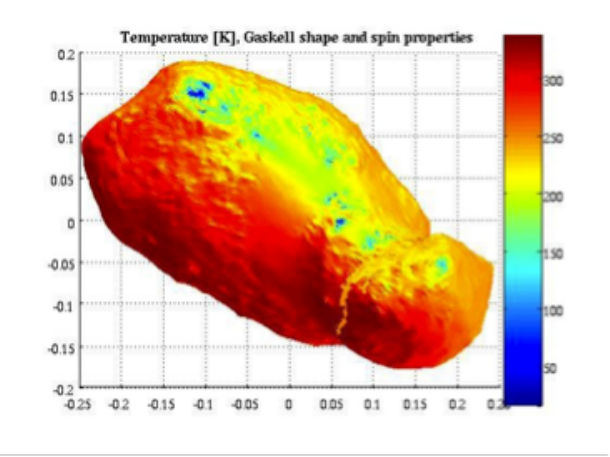}}
  \caption{Comparison of the shape model for Itokawa with 204 facets (left) from
           lightcurve inversion technique, and a much more detailed shape model
           with 49152 facets (right) based on in-situ measurements from the
           Hayabusa mission. Figure and thermo-physical model adapted from
           M\"uller et al.\ \cite{mueller14a}.
     \label{fig:radiometry}}
\end{figure}

\subsection{Direct imaging techniques}
Direct imaging techniques are related to measurements by large ground or
space telescopes or by using data from interplanetary missions.
The targets are spatially resolved in the obtained images.
Size and shape information can then be extracted directly.
Direct, accurate measurements of asteroid physical properties are, meanwhile,
possible for the largest several hundreds asteroids. They can be
spatially resolved using the Hubble Space Telescope (HST) or large
ground-based telescopes equipped with adaptive optics (AO)
on the world's largest telescopes
(Keck\footnote{W.\ M.\ Keck Observatory AO systems: see {\tt https://www2.keck.hawaii.edu/\-optics/\-ao/}},
VLT\footnote{More information about the AO systems of the Very Large Telescope of the
European Southern Observatory ESO can be found at {\tt http://www.eso.org/\-sci/\-facilities/\-develop/\-ao/\-sys.html}},
and Gemini\footnote{More information about the AO systems of the Gemini Observatory 
is given at {\tt http://www.gemini.edu/\-sciops/\-instruments/\-adaptive-optics}}).
The AO systems today are capable of providing images close to the diffraction
limit of the telescope at shorter wavelengths ($<$1.6\,$\mu$m), hence
with an angular resolution of $\approx$33\,milli-arcsecond (mas). Combining this 
technique with lightcurve inversion modelling it is possible to derive
the volume-equivalent diameters for asteroids with typical uncertainties
lower than 10\%, caused by both the uncertainty in the size of the AO
contour and the convex shape model imperfections. It can also remove the
inherent mirror pole ambiguity of lightcurve inversion models
(Marchis et al.\ \cite{marchis06}; Hanu\v{s} et al.\ \cite{hanus13a}).
AO techniques are also capable of discovering binary systems which are important for
studies on internal structure and composition through density determinations
(via mass determination from Kepler's 3$^{rd}$ law; e.g., Merline et al.\
\cite{merline02}; Marchis et al.\ \cite{marchis05}; Descamps et al.\ \cite{descamps07}).

\section{Targets and scientific questions}
\label{sec:targets}

For our benchmark study on minor bodies we selected important targets
which were already visited by spacecraft (or will be visited soon),
which have a wealth of data from different observing techniques available
(or are candidates for being observed with new techniques), but also those
which are or will be useful in the calibration context, or which will allow us
to address or answer specific scientific questions.
The target list is not completely fixed and new objects
--like the Halloween NEA 2015~TB$_{145}$ (M\"uller et al.\ \cite{mueller17a})--
can be included, if we see great observing and/or modelling prospects.
Other targets might
eventually be set aside for lack of data or data quality reasons.

\subsection{Near-Earth asteroids}

Several NEAs have already been visited by interplanetary
missions. We selected the following objects:
\begin{itemize}
\item (433) Eros (visited by NEAR-Shoemaker mission)
\item (25143) Itokawa (visited by Hayabusa mission)
\item (162173) Ryugu (Hayabusa-2 mission; arrival in 2018)
\item (101955) Bennu (OSIRIS-REx mission; arrival in 2018)
\item 2015 TB$_{145}$ ("Halloween 2015 asteroid"; near-Earth flyby)
\end{itemize}

The selection was done on the basis of available visual and thermal
data and the knowledge of object properties from in-situ studies.
The objects have a wealth of visual lightcurve observations available,
allowing the determination of shape and spin information via
lightcurve inversion techniques. For Eros and Itokawa successful 
radar measurements have been conducted. All NEAs have thermal
measurements from different origin, covering wide wavelength
and phase angle ranges.

The work on these targets will help us
to verify the reconstruction of shapes (convex and non-convex solutions)
from multi-apparition lightcurves,
in direct comparison with true object shape and spin properties from
in-situ measurements. Radiometric models "translate" infrared fluxes
into size and albedo solutions, but how accurate are these radiometric
sizes? Is it possible to constrain the spin-axis orientation
from thermal measurements?
Another key science goal is related to the influence of thermal
inertia and surface roughness on the observed infrared fluxes (see
for example Keihm et al.\ \cite{keihm15}). Are there
ways to break the degeneracy between these two properties which both
influence the surface temperatures and the temperature distribution?
Extensive work will be conducted on (162173) Ryugu, the Hayabusa-2 mission
target: a pre-encounter characterisation of this small body will be
used as a reference design model for the planning of the Hayabusa-2 close-distance operation,
and in suppport of the interpretation of in-situ measurements.
Our plans for Bennu, the OSIRIS-REx mission target, are restricted to 
independent tests of the current object properties specified in
the Design Reference Asteroid document (Hergenrother et al.\ \cite{hergenrother14})
by applying our new SBNAF tools.

\subsection{Main-belt asteroids}

For the main-belt asteroids we also selected a few reference objects
with "ground-truth" information from interplanetary missions: Ceres \& Vesta
which were visited by DAWN, Lutetia and Gaspra which were seen during flybys
of Rosetta and Galileo\footnote{A NASA mission which studied Jupiter and its
moons and several other small bodies, see {\tt https://www.jpl.nasa.gov/\-missions/\-galileo/}}
missions, respectively. In addition, we selected a
sample of large main-belt asteroids where the Gaia mission will eventually
provide robust mass estimates (Gaia perturbers). Our goal is to collect existing data (lightcurves,
thermal measurements, occultations, etc.) and conduct new measuremets that will
allow us to derive reliable 3-D shape models and high-quality size estimates. The 
combination of Gaia and SBNAF properties will lead to object densities with
unprecedented accuracy. Another MBA group - partly overlapping with the Gaia mass targets - 
is the "calibrator sample". Selected large and well-known main-belt asteroids are
useful celestial calibrators for many ground-based and space projects, mainly
at far-IR, submm, and mm wavelengths. For the calibration
aspect it is usually necessary to observe bright, point-like sources where flux
predictions are reliable. Direct applications are: absolute flux calibration for
photometers, determination of relative spectral response functions, characterisation
of instrument linearities, testing for filter leaks, or verification of satellite
pointing and tracking capabilities. In addition, we added (911) Agamemnon as 
one of the very few better-known Trojan asteroids. 

\begin{itemize}
\item (1) Ceres \& (4) Vesta (visited by the DAWN mission)
\item (21) Lutetia (Rosetta flyby)
\item (951) Gaspra (Galileo flyby)
\item Gaia mass sample (or Gaia perturbers)
\item Calibrator sample (for thermal IR, submm, mm projects)
\item (911) Agamemnon (unique Trojan asteroid)
\end{itemize}

The main science goals are similar to the ones for NEAs, but now for much larger
objects, for low-conductivity surfaces covered by fine-grain regolith, more spherical shapes,
and for targets of very different taxonomic types. Testing of radiometric solutions
and shape-spin properties is as important as the support for Gaia and for 
calibration applications. The work on main-belt asteroids will also allow us to
study subsurface emission at submm and mm wavelengths
(data coming from ALMA, APEX, IRAM, but also Herschel and Planck).

\subsection{Centaurs and trans-Neptunian objects}

New tools have to be developed to characterise these very remote bodies.
The distant objects can only
be seen under small phase angles (and always very similar aspect angles)
and standard lightcurve inversion techniques fail.
Here, the lightcurves need to be combined
with information from infrared data and occultation measurements to
constrain physical and rotational properties. Thermal data became available
recently (mainly from Spitzer and Herschel, partly also from ALMA and WISE),
and still await a full scientific exploitation. Occultation measurements 
- however - are very difficult due to significant uncertainties
in astrometric positions of the stars and - even more critical - of the
distant solar system objects themselves.
In this distant-object category we focus on large and 'observable' objects:

\begin{itemize}
\item Centaurs with thermal measurements (WISE, Spitzer, Herschel, others)
\item large TNOs with thermal measurements (Spitzer, Herschel, ALMA)
\item Centaurs/TNOs with stellar occultation information
\item all dwarf planets and large dwarf planet candidates within the TNO population
\end{itemize}

The TNO/Centaur target sample is mainly driven by successful occultation 
observations, but recently we also obtained very
high quality and long-coverage lightcurve measurements via the
Kepler-K2 mission. These observations put strong constraints
on object properties, and in combination with the thermal
data will lead to significant improvements in the characterisation
of Centaurs and TNOs. It will also be very interesting to phase in
new ALMA observations at mm wavelengths: here we start to
see subsurface emission telling us details about the top-layer
surface properties, and related emissivity effects such as were seen
with Herschel/PACS/SPIRE for a small sample of TNOs/Centaurs
(Fornasier et al.\ \cite{fornasier13}).

\section{Science results: synergies from ground and space}
\label{sec:synergies}

Remote observations and in-situ measurements of asteroids are considered
highly complementary in nature: remote sensing shows the global picture, but transforming
measured fluxes in physical quantities frequently depends upon model
assumptions to describe surface properties. In-situ techniques measure
physical quantities, such as size, shape, rotational
properties, geometric albedo or surface details, in a more direct way.
However, in-situ techniques are often limited in spatial/rotational/aspect coverage
(flybys) and wavelength coverage (mainly visual and near-IR wavelengths).
Mission targets are therefore important for comparing properties
derived from disk-integrated measurements
with those produced as output of the in-situ measurements.
The associated benefits are:
(i) the model techniques and output accuracies for remote, disk-integrated
observations can be validated (e.g., M\"uller et al.\ \cite{mueller14a} for
the Hayabusa mission target (25143)~Itokawa; O'Rourke et al.\ \cite{orourke12}
for the Rosetta flyby target (21)~Lutetia); (ii) the improved and validated
model techniques can then be applied
to many similar objects easily accessible by remote observations, but which are not
included in interplanetary mission studies.
The pre-mission observations are also important for determining the object's
thermal and physical conditions in support for the construction
of the spacecraft and its instruments, and to prepare flyby, orbiting and
landing scenarios.

We summarise a few key points arising from the combination of
ground and space measurements:
\begin{itemize}
\item disk-resolved vs.\ disk-integrated measurements
\item different viewing geometries from ground and space
\item long-term observation possibilities from ground vs.\ short single flyby events
\item extended (infrared) wavelength coverage from space
\item higher sensitivity from space (or close-up observations)
\item long-term stable observations from space without weather, day/night or airmass issues
\item testing of thermal model parameters (like surface roughness) vs.\ in-situ properties
\item testing of mathematical shape/spin model solutions vs.\ in-situ properties
\item assessment of errors, biases, limitations, etc.\ related to lightcurve inversion,
      radar, occultation, AO, or radiometric techniques
\end{itemize}

In the following projects we took advantage of
the synergetic effect of combining observations from very
different sources.

\subsection{Thermal measurements from Herschel \& Spitzer combined with occultation information}

The multi-chord stellar occultation by the trans-Neptunian object (229762) 2007 UK$_{126}$
led to an area-equivalent radius (of an ellipse fit) of 319$^{+14}_{-7}$\,km,
a geometric V-band albedo of 0.159$^{+0.07}_{-0.013}$ and a possible body oblateness
close to 0.1 (Benedetti-Rossi et al.\ \cite{benedetti16}). The combination of
occultation results with thermal data from the Herschel Space Observatory
allowed us to put additional constraints on the size, shape, possible spin-axis 
orientation, and thermal inertia ($\Gamma$ = 2 - 3\,Jm$^{-2}$s$^{-0.5}$K$^{-}$)
of the main body, but also some upper limits
on the satellite of 2007 UK$_{126}$ (Schindler et al.\ \cite{schindler17}).

A similar approach was possible for the Plutino 2003 AZ$_{84}$
(Santos-Sanz et al.\ \cite{santossanz17}). The Herschel and Spitzer
measurements, in combination with a successful occultation event,
gave a consistent solution for the object's size and shape (close to
a sphere) and favours a close to pole-on orientation of the spin axis.
A new study combined four different occultation events for 2003 AZ$_{84}$
(Dias-Oliveira et al.\ \cite{dias17}) and came to the same conclusion
that we are currently seeing the object close to pole-on. This is also consistent with
the small amplitude of the optical lightcurve (around 0.07\,mag).

A publication on the Eris-Dysnomia system as seen with the synergy of
occultation and thermal emission measurements is currently in preparation
(Kiss et al., in preparation). The results of the first successful 
occultation measurements of Haumea (from January 21, 2017) will
also be combined with the available thermal measurements from 
Herschel and Spitzer (Ortiz et al., in prep.).

\begin{figure}[h!tb]
 \resizebox{\hsize}{!}{\includegraphics{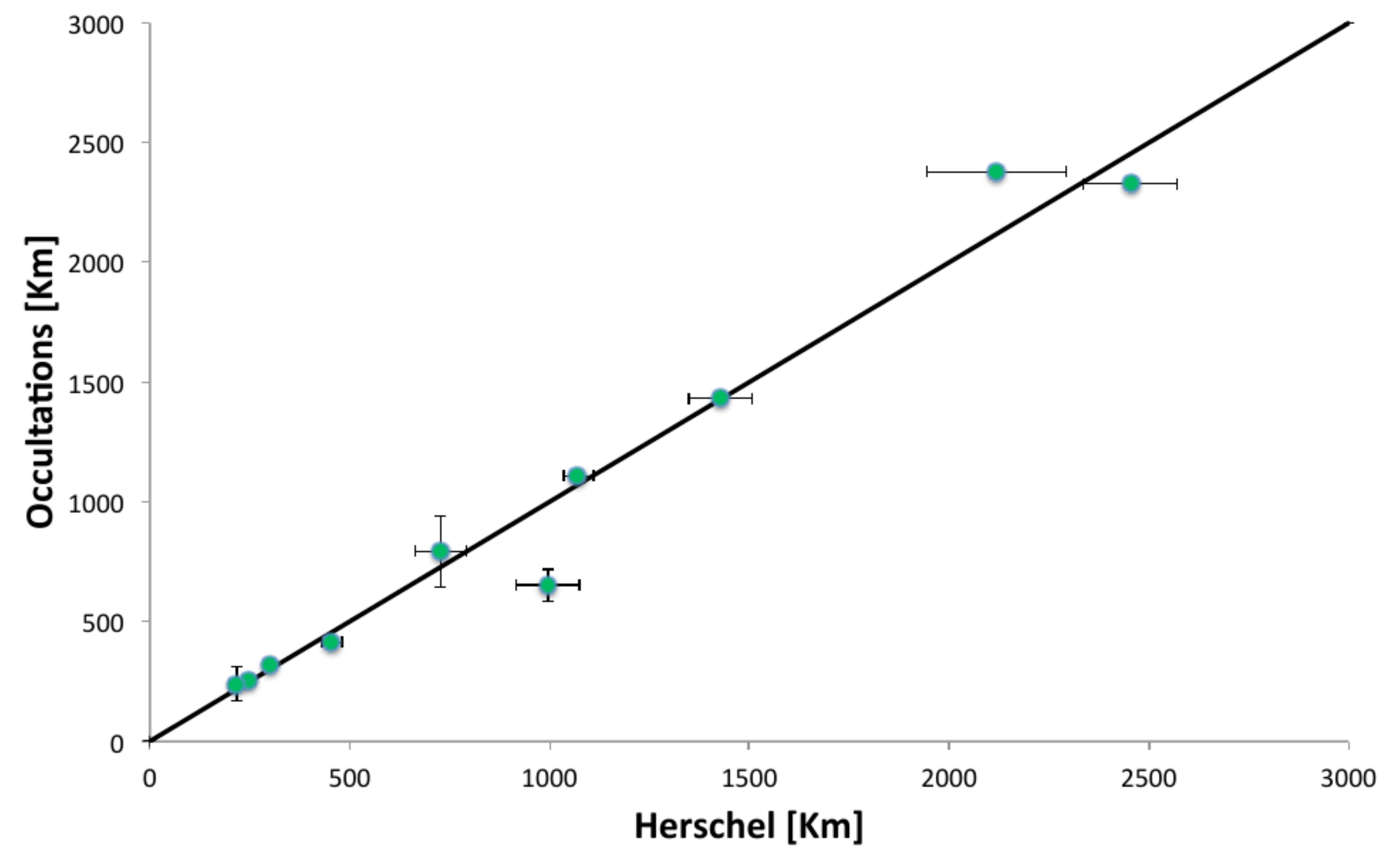}}
  \caption{A comparison between TNO sizes derived from successful
           occultation events and radiometric sizes (mainly related
           to the "TNOs-are-Cool" Herschel key project; M\"uller et al.\
           \cite{mueller09}). Ourliers are
           explained in the text.
     \label{fig:occ_vs_radiom}}
\end{figure}

Figure~\ref{fig:occ_vs_radiom} shows a comparison between size determinations
based on radiometric techniques (mainly based on Herschel thermal measurements)
and occultation measurements. There is a very good agreement between both
completely independent solutions. The point out of the line near 1000\,km is Sedna
that has so-far only a single chord occultation. In case of Pluto (second outlier
with the largest occultation size in Fig.~\ref{fig:occ_vs_radiom})
the (standard) radiometric method is uncertain.
The atmosphere, strong albedo variations and the large satellite Charon would have
to be considered to obtain a reliable radiometric size.

Ideas on how to push the field of TNO occultations are
presented in Santos-Sanz et al.\ (\cite{santossanz16}).
In the near-term future JWST will provide opportunities to
observe stellar occultations by TNOs (and other objects)
with unprecedented accuracy. At the same time, JWST also
has the option to measure the thermal emission up to 28\,$\mu$m
for many small bodies, including the largest TNOs and Centaurs.

\subsection{Kepler-K2 lightcurves combined with thermal data}

The irregular Neptune satellite Nereid was seen in a multiple-day
observing campaign of the Kepler-K2 mission. The results were
presented in a publication "Nereid from space: rotation, size and shape
analysis from K2, Herschel and Spitzer observations" by Kiss et al.\ (\cite{kiss16}).

The Kepler-K2 fields covered also more than 50 Trojan asteroids
(Szab\'o G.\ et al.\ \cite{szabo17}) and a large number of MBAs
(Szab\'o R.\ et al.\ \cite{szabo16}). Both samples comprise a fantastic
laboratory for combined lightcurve-thermal studies. The Kepler-K2
lightcurves - sometimes complemented by auxiliary lightcurve measurements
from ground - constrain the shape and spin properties of these objects.
The thermal measurements (coming from IRAS, MSX, AKARI, WISE, and others)
allow the scaling of the shape solutions, and lead to albedo and thermal
characteristics for each object (this work is planned as part of the SBNAF
project).

Another successful example of Kepler-K2 and Herschel synergies was
presented by P\'al et al.\ (\cite{pal16}): "Large Size and Slow Rotation of the Trans-Neptunian
Object (225088) 2007 OR$_{10}$ Discovered from Herschel and K2 Observations".
2007 OR$_{10}$ is now considered as the "Largest Unnamed World in the Solar
System" (NASA JPL News May 11, 2016). A deeper look at HST archive images
of the same target revealed a small satellite (Marton et al.\ \cite{marton16};
Kiss et al.\ \cite{kiss17}). Follow-up studies with HST are in preparation.

\subsection{Visible and thermal photometric studies}

Does the centaur (54598) Bienor have a ring system? Lightcurve
studies over more than 15 years show a strong decline in amplitude
and, combined with the spectroscopic detection of water ice, this
would be best explained by rings (Fern\'andez-Valenzuela et al.\
\cite{fernandez17}). The main body seems to have an extreme shape
approximated by a triaxial Jacobi ellipsoid, and a possible
density between 0.6 and 0.7\,gcm$^{-3}$. Thermal measurements from Herschel 
(Duffard et al.\ \cite{duffard14}) put
constraints on size and albedo. Different spin-pole orientations
are dicussed in the light of visible and thermal measurements,
but follow-up observations or a high-quality stellar occultation
are needed to clarify the true nature of this exotic centaur.

Double-peaked visible lightcurves indicate that 2008~OG$_{19}$ is a highly
elongated TNO (Fern\'andez-Valenzuela et al.\ \cite{fernandez16}).
The object has a rotation period of about 8.7\,h and
a large peak-to-valley lightcurve amplitude of 0.44\,mag.
A size estimate of 619$^{+56}_{-113}$\,km
is related to an average albedo for scattered disk objects
(Santos-Sanz et al.\ \cite{santossanz12}).
Assuming the object to be in hydrostatic equilibrium gives
a lower limit for its bulk density of just above 0.5 gcm$^{-3}$.
It belongs to the dwarf planet candidates and seems to resemble
the strangely shaped classical Kuiper Belt object Varuna. As for other
large TNOs, the interpretation of visible lightcurves benefits
from having thermal measurements which are less affected by
albedo variations on the surface (see other TNO cases mentioned
before). In the near-term future, it is expected that JWST will
conduct thermal emission studies of selected and scientifically
interesting TNOs like 2008~OG$_{19}$.

For very remote objects standard lightcurve inversion
techniques fail to produce shape and spin solutions due to
the small changes in aspect angles (as seen from a ground-based
or near-Earth observing point). However, the rotational
variability can tell us many things about these large and 
often icy bodies. Santos-Sanz et al.\ (\cite{santossanz17})
obtained thermal lightcurves of the dwarf planet Haumea, and the Plutinos 2003 VS$_2$
and 2003 AZ$_{84}$ with Herschel Space Observatory-PACS. In addition to
size, albedo and shape constraints, the authors also found very low
thermal inertias which seem to be explained best by a moderately
porous regolith where the sub-cm-sized amorphous ice grains have
only loose contacts. Since crystalline water ice signatures are
seen for Haumea, the best explanation points to the presence of
amorphous ice at cm depths below a thin layer of transparent crystalline ice.

Recently, M\"uller et al.\ (\cite{mueller17b}) and Perna et al.\ (\cite{perna17})
presented a study on the
NEA 162173 Ryugu, the target for the Hayabusa-2 mission. The almost 
spherical shape of the target together with the insufficient lightcurve 
quality required the application of radiometric and lightcurve inversion 
techniques in different ways to find the object's spin-axis orientation, 
its shape and to improve the quality of the key physical and thermal 
parameters. The work also showed the difficulties and problems occuring
in combined visible and thermal photometric studies. The constraints
on object properties strongly depend on the quality of the visible
lightcurves, the observing/illumination geometries and wavelengths of the
thermal measurements, and also on the validity of model concepts.

The Spitzer/Herschel sample contains about 30 multiple TNO systems.
The satellite orbits of many of these binary and triple TNOs are sufficient to
derive a system mass. The combination of the masses with volume estimates 
(coming either from radiometric or occultation sizes) determines the bulk density
(e.g., Santos-Sanz et al.\ \cite{santossanz12}; Mommert et al.\ \cite{mommert12}; 
Vilenius et al.\ \cite{vilenius12}).
Kovalenko et al.\ (\cite{kovalenko17}) presented densities ranging from
below 0.5 to almost 4.0\,g\,cm$^{-3}$, with a moderately strong correlation between
diameter and bulk density. Multiple TNOs also strongly correlate with heliocentric
orbital inclination and with magnitude difference between components of multiple system.
Single and multiple TNOs also show different size distributions, but here the
small Centaurs (mainly single TNOs) and the large dwarf planets (almost all are 
multiple systems) cause a bias in the distributions. This work is also a 
nice example for the interplay between different techniques: high-resolution/AO imaging (HST and
large ground-based telescopes) is needed to discover and characterise satellites, while thermal measurements
and occultations are key to constraining the size and volume of these bodies.

The New Horizons mission provided a very detailed view
of the Pluto-Charon system. But only through thermal measurements
with Herschel's long-wavelength channels it was possible to see a strongly
decreasing brightness temperature towards submm wavelengths (Lellouch et 
al.\ \cite{lellouch16}). The best explanation is a spectral emissivity that
decreases steadily from 1 at 20-25\,$\mu$m to $\approx$0.7 at 500\,$\mu$m,
similar to what is found for other icy bodies in the solar system
(Fornasier et al.\ \cite{fornasier13}).
The effect is likely related to a transparent top-layer surface combined
with a strong dielectric constant.

\subsection{Support for interplanetary missions}
The JAXA Hayabusa-2 mission was approved in 2010 and launched on December 3, 2014.
The spacecraft will arrive at the NEA 162173 Ryugu (1999 JU$_3$) in 2018 where it will
perform a survey, land and obtain surface material, then depart in December 2019
and return to Earth in December 2020. We support the mission by providing key
physical, rotational, thermal, and compositional properties of the mission target.
Based on all available visible lightcurve observations and thermal emission
measurements, we applied lightcurve inversion and radiometric techniques. The 
summary of our work (as of late 2016) is documented and discussed in
M\"uller et al.\ (\cite{mueller17b}). Our results are part of the Ryugu
reference model which is widely used in the planning of mission operation
scenarios, the satellite approach phase, touch-down simulations, landing site
selection procedures, and also for outreach activities.
The pre-encounter characterisation will also help to adjust temperature- and
brightness-critical instrument settings. Ryugu will be one of the very few
targets were preencounter properties - derived from disk-integrated remote
observations - can be directly compared to in-situ properties. This will allow
us to verify our model procedures and to consolidate the related error estimates.

A study on the NASA OSIRIS-REx mission target (101\,955) Bennu 
(M\"uller et al.\ \cite{mueller12}) contributed to the official 
"Design Reference Asteroid for the OSIRIS-REx Mission Target (101955) Bennu"
(Hergenrother et al.\ \cite{hergenrother14}). We plan to apply our SBNAF tools
(lightcurve inversion, combination with radar and thermal measurements) to
the available observational data to derive physical and thermal properties
(and their uncertainties) for a direct comparison with in-situ findings
(arrival of OSIRIS-REx at Bennu will be around September 2018).

A new field for the application of our tools and expertise is related
to the topic of in-space utilisation of asteroids. In preparation of 
interplanetary missions with the goals of asteroid mining and
resource utilisation for long-term missions and manned space exploration,
it is necessary (i) to find suitable objects; (ii) to characterize the 
global objects properties (size, shape, rotation) and rough composition;
(iii) to investigate surface and sub-surface characteristics
(Graps et al.\ \cite{graps16}; M\"uller et al.\ \cite{mueller17a}). The SBNAF
project will contribute in different ways to the characterization of these
small NEAs, in close contact to asteroid mining companies.

\section{Tools and services for small bodies}
\label{sec:results}

\subsection{SBNAF public website}

The purpose of the SBNAF public website\footnote{\tt http://www.mpe.mpg.de/\-{\textasciitilde}tmueller/\-sbnaf/}
is to share our scientific interest on small bodies
with the astronomical community and the interested public.
We document the progress and knowledge of the SBNAF project.
We advertise outreach activities that we and other institutions around
Europe organise. Ultimately, we hope to stir curiosity and 
to provoke questions about these rocky and icy bodies,
some of which were around already during the formation of
the planets in our solar system.

On our public website we
present the basic facts about the SBNAF project,
target lists, observing techniques, and all
results (web services, tools, products, databases, predictions, etc.)
for the general planetary science community. We also list all
SBNAF-related publications, conference contributions, and outreach
activities.

\subsection{ISAM}
The Interactive service for asteroid model visualisation (ISAM) is
a web-based service where all kinds of asteroid shape models are
provided, together with various tools to visualise shape solutions,
illumination geometries, rotational properties, disk-integrated
lightcurves, etc.
The SBNAF-related spin and shape models will be provided to the
community via the ISAM service\footnote{\tt http://isam.astro.amu.edu.pl/},
typically in the context of SBNAF project deliverables or after specific
publications. Many of these shape and spin solutions are also used
for other work packages (like for the Gaia or calibration samples).
In the last stage of the project, the full models solutions (convex and non-convex
shape solutions) with absolute sizes and model uncertainties will be published
and provided to the scientific community by the ISAM service.
This service is also of great help for a direct comparison between
on-sky projected shape model solutions with results from an occultation
event or from single-epoch AO images. The ISAM lightcurve prediction for
a given shape and spin solution can be directly compared with lightcurve
measurements or archive data (e.g., from the CDS\footnote{Centre de Donn\'ees
astronomiques de Strasbourg, see {\tt http://cdsweb.u-strasbg.fr/}}
or from LCDB).

\subsection{Gaia-GOSA}
Gaia-Groundbased Observational Service for Asteroids (Gaia-GOSA\footnote{\tt http://www.gaiagosa.eu/};
Santana-Ros et al.\ \cite{santanaros14}; \cite{santanaros16})
is an interactive tool aiming to facilitate asteroid observers in contributing
to the Gaia mission (European Space Agency) by gathering lightcurves of
selected targets. GOSA users can plan their observing runs by selecting the
visible targets for a given date, collaborate with other observers and upload
the frames obtained.
We are responsible for analyzing the data, publishing the results in the
website and creating a lightcurve catalog. Once calibrated, lightcurves will
be easily included in the analysis of Gaia data, which will allow to enhance
the determination of asteroids' physical properties. This service is meanwhile
used by a large community of observers to plan, conduct and provide 
dedicated asteroids measurements.

\subsection{Calibration project}
Celestial standards play a major role in observational astrophysics.
They are needed to characterize the performance of instruments and are
paramount for photometric calibration.    
Large main-belt asteroids fill the flux gap between the submm
calibrators Mars, Uranus and Neptune, and the mid-IR bright calibration
stars. Space instruments at thermal infrared (IR) wavelengths use asteroids
for various calibration purposes: pointing tests, absolute flux calibration,
determination of the relative spectral response function, observing mode
validation, cross-calibration aspects, and several other aspects where bright,
point-like, easily accessible, and reliable sources are needed (M\"uller \& Lagerros
\cite{mueller02}; M\"uller et al.\ \cite{mueller14b}).

The SBNAF project supports worldwide calibration activities for
ground-/airborne-/balloonborne-/space-projects at mid-IR/far-IR/submm/mm wavelengths
by providing highly reliable model predictions of selected well-known asteroids.
These activities are documented in a series of deliverables which are produced
as part of Work Package 4\footnote{\tt http://www.mpe.mpg.de/\-{\textasciitilde}tmueller/\-sbnaf/\-results/\-WP4\_AstCal.html}
of the SBNAF project. The documents explain the
calibration need, the reasons for using asteroids in the calibration context,
and the various steps from simple model predictions for calibration planning
purposes, up to the establishment of new primary calibrators for highly
demanding applications. The asteroid-calibration activities are related to
all infrared space projects (IRAS, MSX, Spitzer, AKARI, WISE, Herschel, Planck), but
also to ongoing submm and mm observatories like ALMA, APEX, IRAM,
or LMT, as well as for balloon-borne (BLAST) and airborne (SOFIA) projects.
A recent large delivery of calibration products (1433 FITS files) includes
specific model predictions for all Herschel observations of 28 asteroids (all
large MBAs) which were used for various calibration activities for the three
instruments PACS, SPIRE, and HIFI (see Herschel Ancillary Data
Products\footnote{\tt https://www.cosmos.esa.int/\-web/\-herschel/\-ancillary-data-products}).

The calibration goals of projects like ALMA or Herschel are very 
ambitious and the traditional celestial calibration sources are 
often not sufficient, showing far-IR excess emission, variability,
unsuitable flux levels or spectral slopes, or modelling shortcomings.
Also the number of celestial calibrator sources is very small and often
there are no sources available in a given observing campaign. The asteroids
are therefore useful to complement existing calibration schemes. 
Currently, the highest demand is coming from Herschel and AKARI (post-operation calibration
activities), SOFIA, ALMA, and sporadically also from other groundbased
observatories like IRAM or APEX. In close contact with instrument calibration
scientist we provide specific asteroid model predictions together with
a documentation on the model input parameters and model quality issues.
Our SBNAF goal for Herschel and AKARI is to make a quality upgrade for the
already implemented asteroid calibration models.

\subsection{Herschel Science Archive (HSA) user-provided products}

The SBNAF project produces high-quality data products for small body observations
of the Herschel Space Observatory. These products are based on sophisticated,
solar-system specific reduction and calibration procedures. The products cover many
Herschel science projects, as well as calibration observations. 

Most Centaur and trans-Neptunian object observations were performed in the framework
of the "TNOs are Cool!" Herschel Open Time Key Program (M\"uller et al., \cite{mueller09};
\cite{mueller10}),
supplemented by some DDT observations of extreme objects (see e.g.\ Kiss et al., 
\cite{kiss13}; P\'al et al., \cite{pal15}). In these cases the standard
HSA\footnote{\tt http://archives.esac.esa.int/\-hsa/\-whsa/} product generation
reduce the data in the co-moving frame and does not combine multiple
observations of the same target. TNO observations were designed in a way that maps
taken at different epochs can serve as mutual backgrounds and a proper combination
of maps can eliminate the background, leaving the target to be the only notable source
in the combined maps (for details, see Kiss et al., \cite{kiss14}) - background
elimination in the far-infrared is crucial due to the strong confusion noise
(see e.g. Kiss et al., \cite{kiss05}).
The simplest maps of this kind are the {\it differential} maps, with one positive
and one negative beam of the object. Further processing ({\it double-differential} maps)
combines these beams and provides a single beam of the target that is preferably
used for photometry. These double-differential maps gave the best photometric
accuracy among all Herschel/PACS scan maps observations (see e.g.\ the case of comet
Hale-Bopp, Szab\'o et al., \cite{szabo12}). Differential and double-differential
maps combine up to eight observations, depending on the band used.
{\it Supersky-subtracted} maps (Santos-Sanz et al., \cite{santossanz12},
Kiss et al., \cite{kiss14}) are also produced -- in this case a 'target-free',
combined background map is created and subtracted from the individual maps.

Similar techniques could be used in the case of near-Earth asteroids if the
displacement of the target was sufficiently small within a specific scan map
observation block (so-called repetition, see e.g.\ the case of Ryugu,
M\"uller et al., \cite{mueller17b}). In most near-Earth asteroid observation,
however, the scan map data had to be reduced in the co-moving frame due to the
fast apparent motion. In these cases our UPDPs\footnote{User Provided
Data Products} use a combination of
observations different from that of the standard HSA products and are also
reduced with an optimised pipeline, in many cases eliminating sub-blocks
affected by instrumental effects. This optimised pipeline - both for near-Earth
asteroids and TNO observations - include specifically selected high-pass filter,
masking and de-glitching parameters, signal drift correction, and optimally
chosen pixel size and back-projection pixel fraction values (for details, see
Kiss et al., \cite{kiss14}). Our UPDPs are provided as standard FITS files,
accompanied by specific release notes (see HSC
pages on UPDP\footnote{\tt https://www.cosmos.esa.int/\-web/\-herschel/\-user-provided-data-products}).

\subsection{Occultation predictions}
Based on publicly available and own tools, together with multiple campaigns of dedicated
astrometric observations, we make regular predictions of occultation events produced by
our SBNAF sample targets. Step-by-step we also phase in information from the Gaia mission
to improve the accuracy of the predicted shadow paths. We focus on occultation predictions
for Europe (mainly Spain) and South America (mainly Argentina, Chile, and neighbours)
where we have own telescope facilities and close contacts with
networks of well-trained observers. In the near-term future we
might also include potentially interesting asteroid occultations of strong, compact
radio sources, following up on the work by Lehtinen et al.\ (\cite{lehtinen16}).
The predictions of MBA events is, meanwhile, very reliable and easily possible
via public tools (e.g., {\it Occult Watcher}\footnote{\tt http://www.occultwatcher.net/}
or the occultation prediction software
{\it Occult}\footnote{\tt http://www.lunar-occultations.com/\-iota/\-occult4.htm}).
Our focus in the SBNAF project lies on the much more challenging predictions for TNOs and Centaurs. 
These distant objects have a very small projected size in the sky:
$\approx$8\,mas for a typical 100\,km Centaur at 17\,AU, $\approx$82\,mas for
the largest TNO (Pluto), and $\approx$14\,mas for a 400\,km TNO at 39\,AU.
This means that to catch a stellar occultation produced by one of these tiny bodies
we must have a very accurate knowledge of their orbits and ephemeris, and we also
need very precise star catalogues with accurate positions in the
occultation track region.
So far, the most accurate catalogs were the UCAC4 catalog with uncertainties of
$\approx$15-100\,mas (Zacharias et al.\ \cite{zacharias13}) and the (incomplete)
URAT1 catalog with up to $\approx$2-3\,times better astrometric precision
(Zacharias et al.\ \cite{zacharias15}). The Gaia DR1 catalogue (Lindegren et
al.\ \cite{lindegren16}) improves the situation for the star positions
considerably. The uncertainties will be even smaller with future Gaia data releases
(the second one is expected for April 2018\footnote{\tt https://www.cosmos.esa.int/\-web/\-gaia/\-release}).
However, the orbits of the TNOs and Centaurs are only poorly known and
dominate often the prediction uncertainties. These objects need centuries
or even more time to complete one orbit and up to now we have only observed
a very small arc of their orbits. Here, it is often necessary to switch
to relative astrometry with the target and star imaged together in the same 
field of view (FOV) in the months and weeks before a predicted event.
The FOV must be large in order to have a significant number of astrometric
references. In this way, one can achieve a high astrometric precision for
TNOs up to $\approx$20\,mas (Ortiz et al.\ \cite{ortiz11}), enough to predict
the shadow path on Earth within $\approx$500\,km.
This technique has been succesful for $\approx$10 TNOs (see Fig.~\ref{fig:occ_vs_radiom})
and a few Centaurs. Large groups of observers are using these
predictions to plan own observations for supporting these unique
and scientifically important occultation events.

\subsection{Database of calibrated thermal infrared measurements}
One of the SBNAF goals is to produce a database for thermal infrared
observations of small bodies. The SBNAF concept foresees to start with
a database containing all available (published) thermal IR measurements
for our selected samples of solar system targets, first for internal usage,
later with open access to the public. The database should also have the
future option (after the SBNAF project end) to include large amounts
of thermal data for all solar system small bodies which have been
detected at thermal IR wavelengths.

Our sample includes selected near-Earth, main-belt, trojans, Centaurs,
and trans-Neptunian objects.
All targets have significant amounts of thermal measurements from different
satellite missions (IRAS, MSX, ISO, AKARI, Spitzer, Herschel, Wise, NEOWise),
from SOFIA, and from ground. We assume that all objects appear as point
sources (or point-like sources) and that the photometric measurements are
corrected for aperture losses and well calibrated. For each dataset we will
also give a short recipe on how to use the calibrated in-band fluxes in the
context of radiometric techniques. The recipes include information about the
instrument and calibration aspects, as well as the filter pass bands,
saturation, non-linearity issues, and colour-correction procedures.
The collection of infrared measurements in a unified way has the goal to better
describe these targets by using all available data simultaneously. In this way,
we also want to address a range of scientific questions (see Section~\ref{sec:targets}).

For storing IR measurements in a database, we first focus on the essential basic
entries: object identifier, observatory, measurement identifier,
instrument/-band/filter/mode, start time of the observations, duration,
measured in-band flux (calibrated, aperture-/beam-corrected,
non-linearity/saturation-corrected, etc.) and the corresponding flux error.
A list of quality comments or other flags should also be possible. Some
of the database entries could also have a link to auxiliary information,
like project-/instrument pages, specific archives,
relevant publications and documents.

In addition to the measured flux and error, it is needed to calculate also
the mono-chromatic flux density at pre-defined reference wavelengths,
to translate times from calendar to JD (with/without correction for
light-travel time), to convert wavelengths to frequencies, to add absolute
flux errors (if not done already), etc. For each measurement, it would also
be useful to store (or calculate) information like the heliocentric,
observer-centric distances, and the phase angle, or also orbital parameters.
Optional parameters are RA, Dec, ObsEcLon, ObsEcLat, solar elongation,
ecliptic helio-centric XYZ of the observer and the target, light-travel time,
apparent motion of the object (all at observation mid-time). This could for
example be done via the Callhorizons Python
module\footnote{\tt https://pypi.python.org/\-pypi/\-CALLHORISONS}.
It is planned to make these tools
available together with the first database prototype.
This database will gain importance in the context of planning and analysing
JWST mid-IR measurements of small bodies.

\subsection{Phase curve calculator \& H-magnitude}

When absolute photometry (placed on a standard photometric system) is
available asteroid scattering properties related to the so-called opposition
effect (caused by the backscattering mechanism) can be determined. This sort
of observations typically involve photometric sky conditions and/or the
presence of photometric standards in the field-of-view of the telescope. 
For accurate phase curves full lightcurves obtained every few degrees in phase
angle are needed to determine the magnitude shifts between the measurements
for a range of phase angles. Data obtained at various oppositions require
aspect corrections and sparse data require even more calibrations to correct
for lightcurve amplitude. Once the data have been properly calibrated we will
fit the data using different fitting schemes depending on the quality of the data.
The currently recommended IAU functions are the H, G1, G2 and H, G12 phase
functions (Muinonen et al.\ \cite{muinonen10}).
A variations of the H, G12, can be utilized for low quality data or when the
taxonomic type of the fitted object is known. For low quality (magnitude
uncertanities larger then $\sim 0.3$\,mag), but numerous data it is recommended
to perform the fitting in magnitude space (Penttila et al.\ \cite{penttila16})
to avoid systematic biases. For high quality data the fitting can be done either
in flux or magnitude space. In magnitude space the problem reduces to a linear
one thus reducing the computational time. The uncertanities in phase curve
parametes can be non-Gaussian, thus procedures like for example the bootstraping
method are recommended to estimate the uncertanity.
The scattering properties relate to regolith properties such as spectral type,
albedo, porosity, and others (Oszkiewicz et al.\ \cite{oszkiewicz11}; \cite{oszkiewicz12},
Shevchenko et al.\ \cite{shevchenko16}) .Those relations are currently poorly
determined and understood. Obtaining high number of accurate phase curves is
therefore crucial in establishing those relationships and their implications
to asteroid population as a whole. As part of SBNAF, we will document the
procedures for the general users, and re-calculate phase curves and H-magnitudes
(also very relevant for albedo calculations) for our sample objects. We will
also address critical questions related to phase curves for objects with
albedo variations, multiple objects, or Centaurs/TNOs with potential ring system.
Phase curves and H-magnitudes are crucial for the radiometric analysis of
thermal data (e.g., M\"uller et al.\ \cite{mueller17a} and references therein).

\subsection{Others}
It is also foreseen to develop various new tools, techniques, and services
related to the above mentioned observing techniques, prediction tools and
models. Our database(s) will also be made public towards the end of the
SBNAF project in 2019. One possible long-term option is to merge our database(s) 
with existing services of the "Virtual European Solar and Planetary
Access (VESPA\footnote{Virtual European Solar and Planetary Access database,
see {\tt http://www.europlanet-vespa.eu/}})".
The wealth of new lightcurve data obtained during SBNAF project will be
made available to the public using widely used services like Strasbourg
CDS and ALCDEF, typically together
with a scientific publication. In some cases the data will be shared
with other projects, like the recetly accepted large proposal that plans to
observe 38 large ($>$ 100\,km) MBAs with VLT/SPHERE\footnote{SPHERE is the
extreme AO system and coronagraphic facility at the VLT;
{\tt https://www.eso.org/sci/\-facilities/\-paranal/\-instruments/\-sphere.html}}
in high resolution (see Marsset et al.\ \cite{marsset17}).
Such large bodies are seen as primordial remnants of the early Solar System
(Morbidelli et al.\ \cite{morbidelli09}). The shape, size and spin
reconstruction for these objects will benefit from the combination of
Adaptive Optics images (from VLT/SPHERE), standard lightcurves, occultation,
and thermal information. The derived volume (via their 3-D shape) together
with mass estimates (mainly from Gaia) will allow us to determine their bulk
density and hence to characterize their internal structure. This fundamental
property is very relevant for addressing the questions regarding the earliest
stages of planetesimal formation and their subsequent collisional and
dynamical evolution.

%
%

\section{Outlook} \label{sec:outlook}

The core of the SBNAF project is to study selected small bodies at very 
different distances from the sun. We work on combining 
different observations and modelling techniques, knowing that
very different approaches are required for NEAs, MBAs, Centaurs, or TNOs.
Often we have to complement the analysis by pushing for more
observations. A large and very important element of SBNAF is therefore
the planning and conduction of measurements.
At the same time, we develop new tools and services to combine
information from ground and space, and also from very different
observing techniques. It requires to extract from each method
the constraints for the derived object properties and also the
reliability of these values. By focusing on objects with 
"ground truth" we will put the new tools on solid grounds. Our
results are already partly available from our public webpage\footnote{\tt http://www.mpe.mpg.de/\-{\textasciitilde}tmueller/\-sbnaf/}
and documented in more than 20 peer-reviewed (submitted, accepted and published)
publications.
More scientific publications, new tools, databases and various services
for small-body observers can be expected during our 3-year
project phase (until mid 2019).


\section*{References}

\begin{thebibliography}{}
\bibitem[2017]{alilagoa17}
   Al{\'i}-Lagoa, V., Delbo', M.\ 2017,
   Sizes and albedos of Mars-crossing asteroids from WISE/NEOWISE data,
   A\&A 603, A55, 8 pp
\bibitem[2014]{bartczak14}
   Bartczak, P., Michalowski, T., Santana-Ros, T., Dudzinski, G.\ 2014,
   A new non-convex model of the binary asteroid 90 Antiope obtained with the SAGE modelling technique,
   MNRAS 443, 1802--1809
\bibitem[2014a]{bartczak14a}
   Bartczak, P., Santana-Ros, T., Michalowski, T.\ 2014,
   Non-convex shape models of asteroids based on photometric observations,
   Asteroids, Comets, Meteors 2014. Proceedings of the conference held
   30 June - 4 July, 2014 in Helsinki, Finland. Edited by K. Muinonen et al., 29B
\bibitem[2017]{bartczak17}
   Bartczak, P., Kryszczy{\'n}ska, A., Dudzi{\'n}ski, G.\ et al.\ 2017,
   A new non-convex model of the binary asteroid (809) Lundia obtained with the SAGE modelling technique,
   MNRAS 471, 941--947
\bibitem[2016]{benedetti16}
   Benedetti-Rossi, G., Sicardy, B., Buie, M.\ W.\ et al.\ 2016,
   Results from the 2014 November 15th multi-chord stellar occultation by the TNO (229762) 2007 UK$_{126}$,
   AJ 152, 156, 11 pp
\bibitem[2014]{braga-ribas14}
   Braga-Ribas, F., Sicardy, B., Ortiz, J.\ L.\ et al.\ 2014,   
   A ring system detected around the Centaur (10199) Chariklo,
   Nature 508, 72--75
\bibitem[2012]{carry12}
   Carry, B., Kaasalainen, M., Merline, W.\ J.\ et al.\ 2012,
   Shape modeling technique KOALA validated by ESA Rosetta at (21) Lutetia,
   P\&SS 66, 200--212
\bibitem[2015]{delbo15}
   Delbo, M., Mueller, M., Emery, J.\ P.\ et al.\ 2015,
   Asteroid Thermophysical Modeling,
   in Asteroids IV, P.\ Michel, F.\ E.\ DeMeo, W.\ F.\ Bottke (eds.),
   University of Arizona Press, Tucson, 107--128
\bibitem[2007]{descamps07}
   Descamps, P., Marchis, F., Michalowski, T.\ et al.\ 2007,
   Figure of the double Asteroid 90 Antiope from adaptive optics and lightcurve observations,
   Icarus 187, 482--499
\bibitem[2017]{dias17}
   Dias-Oliveira, A., Sicardy, B., Ortiz, J.-L.\ et al.\ 2017,
   Study of the Plutino Object (208996) 2003 AZ$_{84}$ from Stellar Occultations: Size, Shape, and Topographic Features,
   AJ 154, 22, 13 pp
\bibitem[2009]{duffard09}
   Duffard, R., Ortiz, J.\ L., Thirouin, A.\ et al.\ 2009,
   Transneptunian objects and Centaurs from light curves,
   A\&A 505, 1283--1295
\bibitem[2014]{duffard14}
   Duffard, R., Pinilla-Alonso, N., Santos-Sanz, P.\ et al.\ 2014,
   "TNOs are Cool!": A Survey of the Transneptunian Region. XI. A Herschel-PACS view of 16 Centaurs,
   A\&A 564, A92, 17 pp
\bibitem[2011]{durech11}
   \v{D}urech, J., Kaasalainen, M., Herald, D.\ et al.\ 2011,
   Combining asteroid models derived by lightcurve inversion with asteroidal occultation silhouettes,
   Icarus 214, 652--670
\bibitem[2016]{durech16}
   \v{D}urech, J., Hanu\v{s}, J., Oszkiewicz, D., Vanco, R.\ 2016,
   Asteroid models from the Lowell photometric database,
   2016, A\&A 587, A48, 6 pp
\bibitem[1979]{elliot79}
   Elliot, J.\ L.\ 1979,
   Stellar occultation studies of the solar system,
   ARA\&A 17, 445--475
\bibitem[1992]{elliot92}
   Elliot, J.\ L., Young, L.\ A.\ 1992,
   Analysis of stellar occultation data for planetary atmospheres. I - Model fitting, with application to Pluto,
   AJ 103, 991--1015
\bibitem[2016]{fernandez16}
   Fern{\'a}ndez-Valenzuela, E., Ortiz, J.\ L., Duffard, R.\ et al.\ 2016,
   2008 OG19: a highly elongated Trans-Neptunian object,
   MNRAS 456, 2354--2360
\bibitem[2017]{fernandez17}
   Fern\'andez-Valenzuela, E., Ortiz, J.\ L., Duffard, R.\ et al.\ 2017,
   Physical properties of centaur (54598) Bienor from photometry,
   MNRAS 466, 4147--4158
\bibitem[2013]{fornasier13}
   Fornasier, S., Lellouch, E., M\"uller, T.\ G.\ et al.\ 2013,
   "TNOs are Cool": A survey of the trans-Neptunian region. VIII. Combined
   Herschel PACS and SPIRE observations of nine bright targets at 70–500\,$\mu$m,
   A\&A 555, A15, 22 pp
\bibitem[2016]{graps16}
   Graps, A.\ L., Blondel, P., Bonin, G.\ et al.\ 2016,
   In-space utilisation of asteroids (ASIME 2016 conference White Paper),
   Astrophysics - Earth and Planetary Astrophysics,
   https://arxiv.org/abs/1612.00709, 81 pp
\bibitem[2013]{hanus13}
   Hanu\v{s}, J., \v{D}urech, J., Broz, M.\ et al.\ 2013,
   Asteroids' physical models from combined dense and sparse photometry
   and scaling of the YORP effect by the observed obliquity distribution,
   A\&A 551, A67, 16 pp
\bibitem[2013a]{hanus13a}
   Hanu\v{s}, J., Marchis, F., \v{D}urech, J.\ 2013,
   Sizes of main-belt asteroids by combining shape models and Keck adaptive optics observations,
   Icarus 226, 1045--1057
\bibitem[1998]{harris98}
   Harris, A.\ W.\ 1998,
   A Thermal Model for Near-Earth Asteroids,
   Icarus 131, 291--301
\bibitem[2002]{harris02}
   Harris, A.\ W.\ \& Lagerros, J.\ S.\ V.\ 2002,
   Asteroids in the Thermal Infrared,
   Asteroids III, W. F. Bottke Jr., A. Cellino, P. Paolicchi, and R. P. Binzel (eds),
   University of Arizona Press, Tucson, p.205--218
\bibitem[2014]{hergenrother14}
   Hergenrother, C.\ W., Barucci, M.\ A., Barnouin, O.\ et al.\ 2014,
   The Design Reference Asteroid for the OSIRIS-REx Mission Target (101955) Bennu,
   Astrophysics - Earth and Planetary Astrophysics, https://arxiv.org/abs/1409.4704,
   116 pp
\bibitem[2001a]{kaasalainen01a}
   Kaasalainen, M.\ \& Torppa, J.\ 2001,
   Optimization Methods for Asteroid Lightcurve Inversion. I. Shape Determination,
   Icarus, 153, 24--36
\bibitem[2001b]{kaasalainen01b}
   Kaasalainen, M., Torppa, J.\ \& Muinonen, K.\ 2001,
   Optimization Methods for Asteroid Lightcurve Inversion. II. The Complete Inverse Problem,
   Icarus, 153, 37--51
\bibitem[2015]{keihm15}
   Keihm, S., Tosi, F., Capria, M.\ T.\ et al.\ 2015,
   Separation of thermal inertia and roughness effects from Dawn/VIR
   measurements of Vesta surface temperatures in the vicinity of Marcia Crater,
   Icarus 262, 30--43
\bibitem[2005]{kiss05}
   Kiss, Cs, Klaas, U., Lemke, D.\ 2005,
   Determination of confusion noise for far-infrared measurements,
   A\&A 430, 343--353
\bibitem[2013]{kiss13}
   Kiss, Cs., Szab\'o, Gy., Horner, J.\ et al.\ 2013,
   A portrait of the extreme solar system object 2012 DR30,
   A\&A 555, A3, 13 pp
\bibitem[2014]{kiss14}
   Kiss, C., M\"uller, T.\ G., Vilenius, E.\ et al.\ 2014,
   Optimized Herschel/PACS photometer observing and data reduction strategies for moving solar system targets,
   ExA 37, 161--174
\bibitem[2016]{kiss16}
   Kiss, C., P\'al, A., Farkas-Tak\'acs, A.\ I.\ et al.\ 2016,
   Nereid from space: rotation, size and shape analysis from K2, Herschel and Spitzer observations,
   MNRAS 457, 2908--2917
\bibitem[2017]{kiss17}
   Kiss, Cs., Marton, G., Farkas-Tak\'acs, A.\ et al.\ 2017,
   Discovery of a satellite of the large trans-Neptunian object (225088) 2007~OR$_{10}$,
   ApJL, 838, L1, 5 pp
\bibitem[2017]{kovalenko17}
   Kovalenko, I.\ D., Doressoundiram, A., Lellouch, E.\ et al.\ 2017,
   "TNOs are Cool": A survey of the trans-Neptunian region. XIII. Characterization
   of multiple trans-Neptunian objects observed with Herschel Space Observatory,
   A\&A, accepted, 8 pp
\bibitem[2014]{lacerda14}
   Lacerda, P., Fornasier, S., Lellouch, E.\ et al.\ 2014,
   The albedo-color diversity of trans-Neptunian objects,
   ApJL, 793, L2, 6 pp
\bibitem[1986]{lebofsky86}
   Lebofsky, L.\ A., Sykes, M.\ V., Tedesco, E.\ F.\ et al.\ 1986,
   A refined 'standard' thermal model for asteroids based on observations of 1 Ceres and 2 Pallas,
   Icarus 68, 239--251
\bibitem[2016]{lehtinen16}
   Lehtinen, K., Bach, U., Muinonen, K.\ et al.\ 2016,
   Asteroid sizing by radiogalaxy occultation at 5 GHz,
   ApJL, 822, L21, 5 pp
\bibitem[2013]{lellouch13}
   Lellouch, E., Santos-Sanz, P., Lacerda, P.\ et al.\ 2013,
   "TNOs are Cool!": A Survey of the Transneptunian Region: Thermal properties
   of Kuiper Belt objects and Centaurs from combined Herschel and Spitzer observations,
   A\&A 557, A60, 19 pp
\bibitem[2016]{lellouch16}
   Lellouch, E., Santos-Sanz, P., Fornasier, S.\ et al.\ 2016,
   The long-wavelength thermal emission of the Pluto-Charon system from Herschel
   observations. Evidence for emissivity effects,
   A\&A 588, A2, 15 pp
\bibitem[2016]{lindegren16}
   Lindegren, L., Lammers, U., Bastian, U.\ et al.\ 2016,
   Gaia Data Release 1. Astrometry: one billion positions, two million proper motions and parallaxes,
   A\&A 595, A4, 32 pp
\bibitem[2005]{marchis05}
   Marchis, F., Hestroffer, D., Descamps, P.\ et al.\ 2005,
   Mass and density of Asteroid 121 Hermione from an analysis of its companion orbit,
   Icarus 178, 450--464
\bibitem[2006]{marchis06}
   Marchis, F., Kaasalainen, M., Hom, E.\ F.\ Y.\ et al.\ 2006,
   Shape, size and multiplicity of main-belt asteroids. I. Keck Adaptive Optics survey,
   Icarus 185, 39--63
\bibitem[2012]{marciniak12}
   Marciniak, A., Bartczak, P., Santana-Ros, T.\ et al.\ 2012,
   Photometry and models of selected main belt asteroids. IX. Introducing interactive service for asteroid models (ISAM),
   A\&A 545, A131, 31 pp
\bibitem[2017]{marsset17}
   Marsset, M., Carry, B., Dumas, C.\ et al.\ 2017,
   3D shape of asteroid (6) Hebe from VLT/SPHERE imaging: Implications for the origin of ordinary H chondrites,
   A\&A 604, A64, 12 pp
\bibitem[2016]{marton16}
   Marton, G., Kiss, Cs., M\"uller, T.\ G.\ 2016,
   The moon of the large Kuiper-belt object 2007 OR 10,
   American Astronomical Society, DPS meeting \#48, id.\ 120.22
\bibitem[2002]{merline02}
   Merline W.\ J., Weidenschilling, S.\ J., Durda, D.\ D.\ et al.\ 2002,
   Asteroids Do Have Satellites,
   Asteroids III, W. F. Bottke Jr., A. Cellino, P. Paolicchi, and R. P. Binzel (eds), 
   University of Arizona Press, Tucson, p.289--312
\bibitem[2012]{mommert12}
   Mommert, M., Harris, A.\ W., Kiss, C.\ et al.\ 2012,
   "TNOs are cool": A survey of the trans-Neptunian region
   V. Physical characterization of 18 Plutinos using Herschel-PACS observations,
   A\&A 541, A93, 17 pp
\bibitem[2009]{morbidelli09}
   Morbidelli, A., Bottke, W.\ F., Nesvorny, D., Levison, H.\ F.\ 2009,
   Asteroids were born big,
   Icarus 204, 558--573
\bibitem[2002]{mueller02}
   M\"uller, T.\ G.\ \& Lagerros, J.\ S.\ V.\ 2002,
   Asteroids as calibration standards in the thermal infrared for space observatories,
   A\&A 381, 324--339
\bibitem[2009]{mueller09}
   M\"uller, T.\ G., Lellouch, E., B\"ohnhardt, H.\ et al.\ 2009,
   "TNOs are Cool": A Survey of the Transneptunian Region,
   EM\&P 105, 209--219
\bibitem[2010]{mueller10}
   M\"uller, T.\ G., Lellouch, E., Stansberry, J.\ et al.\ 2010,
   "TNOs are Cool": A survey of the trans-Neptunian region. I. Results from the Herschel science demonstration phase (SDP),
   A\&A 518, L146, 5 pp
\bibitem[2012]{mueller12}
   M\"uller, T.\ G., O'Rourke, L., Barucci, A.\ M.\ et al.\ 2012,
   Physical properties of OSIRIS-REx target asteroid (101955) 1999 RQ36. Derived
   from Herschel, VLT/ VISIR, and Spitzer observations,
   A\&A 548, A36, 9 pp
\bibitem[2014a]{mueller14a}
   M\"uller, T.\ G., Hasegawa, S.\ \& Usui, F.\ 2014,
   (25143) Itokawa: The power of radiometric techniques for the interpretation
   of remote thermal observations in the light of the Hayabusa rendezvous results,
   PASJ 66, 52, 1--17
\bibitem[2014b]{mueller14b}
   M\"uller, T.\ G., Balog, Z., Nielbock, M.\ et al.\ 2014,
   Herschel celestial calibration sources. Four large main-belt asteroids
   as prime flux calibrators for the far-IR/sub-mm range,
   ExA, 37, 253--330
\bibitem[2017a]{mueller17a}
   M\"uller, T.\ G., Marciniak, A., Butkiewicz-B\k{a}k et al.\ 2017a,
   Large Halloween Asteroid at Lunar Distance,
   A\&A 598, A63, 10 pp
\bibitem[2017b]{mueller17b}
   M\"uller, T.\ G., \v{D}urech, J., Ishiguro, M.\ et al.\ 2017b,
   Hayabusa-2 Mission Target Asteroid 162173 Ryugu (1999 JU$_3$):
   Searching for the Object's Spin-Axis Orientation,
   A\&A 599, A103, 25 pp
\bibitem[2010]{muinonen10}
   Muinonen, K., Belskaya, I.\ N., Cellino, A.\ et al.\ 2010,
   A three-parameter magnitude phase function for asteroids,
   Icarus 209, 542--555
\bibitem[2017]{nugent17}
   Nugent, C.\ R., Mainzer, A., Masiero, J.\ et al.\ 2017,
   Observed asteroid surface area in the thermal infrared,
   AJ, 153, id.\ 90, 5 pp
\bibitem[2012]{orourke12}
   O'Rourke, L., M\"uller, T., Valtchanov, I.\ et al.\ 2012,
   Thermal and shape properties of asteroid (21) Lutetia from Herschel
   observations around the Rosetta flyby,
   Planetary and Space Science, 66, 192--199
\bibitem[2011]{ortiz11}
   Ortiz, J.\ L., Cikota, A., Cikota, S.\ et al.\ 2011,
   A mid-term astrometric and photometric study of trans-Neptunian object (90482) Orcus,
   A\&A 525, A31, 12 pp
\bibitem[2012]{ortiz12}
   Ortiz, J.\ L., Sicardy, B., Braga-Ribas, F.\ et al.\ 2012,
   Albedo and atmospheric constraints of dwarf planet Makemake from a stellar occultation,
   Nature 491, 566--569
\bibitem[2015]{ortiz15}
   Ortiz, J.\ L., Duffard, R., Pinilla-Alonso, N.\ et al.\ 2015,
   Possible ring material around centaur (2060) Chiron,
   A\&A 576, A18, 12 pp
\bibitem[2002]{ostro02}
   Ostro, S.\ J., Hudson, R.\ S., Benner, L.\ A.\ M.\ et al.\ 2002,
   Asteroid Radar Astronomy,
   Asteroids III, W.\ F.\ Bottke Jr., A.\ Cellino, P.\ Paolicchi, and R.\ P.\ Binzel (eds),
   University of Arizona Press, Tucson, p.151--168
\bibitem[2011]{oszkiewicz11}
   Oszkiewicz, D.\ A., Muinonen, K., Bowell, E.\ et al.\ 2011,
   Online multi-parameter phase-curve fitting and application to a large corpus of asteroid photometric data,
   Journal of Quantitative Spectroscopy and Radiative Transfer 112, 1919--1929
\bibitem[2012]{oszkiewicz12}
   Oszkiewicz, D.\ A., Bowell, E., Wasserman, L.\ H.\ et al.\ 2012,
   Asteroid taxonomic signatures from photometric phase curves,
   Icarus 219, 283--296
\bibitem[2012]{pal12}
   P\'al, A., Kiss, C., M\"uller, T.\ G.\ et al.\ 2012,
   "TNOs are Cool": A survey of the trans-Neptunian region. VII. Size and
   surface characteristics of (90377) Sedna and 2010 EK$_{139}$,
   A\&A 541, L6, 4 pp
\bibitem[2015]{pal15}
   P\'al, A., Kiss, Cs., Horner, J.\ et al.\ 2015,
   Physical properties of the extreme Centaur and super-comet candidate 2013 AZ$_{60}$,
   A\&A 583, A93, 8 pp
\bibitem[2016]{pal16}
   P\'al, A., Kiss, Cs., M\"uller, T.\ G.\ et al.\ 2016,
   Large Size and Slow Rotation of the Trans-Neptunian Object (225088) 2007 OR$_{10}$
   Discovered from Herschel and K2 Observations,
   AJ 151, id.\ 117, 8 pp
\bibitem[2016]{penttila16}
   Penttil{\"a}, A., Shevchenko, V.\ G., Wilkman, O.\ et al.\ 2016,
   H, G1, G2 photometric phase function extended to low-accuracy data,
   Planetary and Space Science 123, 117--125
\bibitem[2017]{perna17}
   Perna, D., Barucci, A., Ishiguro, M.\ et al.\ 2017,
   Spectral and rotational properties of near-Earth asteroid (162173) Ryugu, target of the Hayabusa-2 sample return mission,
   A\&A 599, L1, 4 pp
\bibitem[2014]{santanaros14}
   Santana-Ros, T., Bartczak, P., Michalowski, T., Tanga, P.\ 2014,
   Gaia-GOSA: An interactive service for asteroid follow-up observations,
   EAS 67-68, 109--112
\bibitem[2016]{santanaros16}
   Santana-Ros, T., Marciniak, A., Bartczak, P.\ 2016,
   Gaia-GOSA: A Collaborative Service for Asteroid Observers,
   The Minor Planet Bulletin 43, 205--207
\bibitem[2012]{santossanz12}
   Santos-Sanz, P., Lellouch, E., Fornasier, S.\ et al.\ 2012,
   "TNOs are Cool": A survey of the trans-Neptunian region. IV. Size/albedo
   characterization of 15 scattered disk and detached objects observed with Herschel-PACS,
   A\&A 541, A92, 18 pp
\bibitem[2016]{santossanz16}
   Santos-Sanz, P., French, R.\ G., Pinilla-Alonso, N.\  et al.\ 2016,
   James Webb Space Telescope Observations of Stellar Occultations by Solar System Bodies and Rings,
   PASP 128, pp 018011
\bibitem[2017]{santossanz17}
   Santos-Sanz, P., Lellouch, E., Groussin, O.\ et al.\ 2017,
   "TNOs are Cool": A Survey of the Transneptunian Region. XII Thermal light curves
   of Haumea, 2003 VS$_2$ and 2003 AZ$_{84}$ with Herschel Space Observatory-PACS,
   A\&A 604, A95, 19 pp
\bibitem[2017]{schindler17}
   Schindler, K., Wolf, J., Bardecker, J.\ et al.\ 2017,
   Results from a triple chord stellar occultation and far-infrared photometry
   of the trans-Neptunian object (229762) 2007 UK$_{126}$,
   A\&A 600, A12, 16 pp
\bibitem[2016]{shevchenko16}
   Shevchenko, V.\ G., Belskaya, I.\ N., Muinonen, K.\ et al.\ 2016,
   Asteroid observations at low phase angles. IV. Average parameters for the new H, G1, G2 magnitude system,
   Planetary and Space Science 123, 101--116
\bibitem[2015]{sonnett15}
   Sonnett, S., Mainzer, A., Grav, T.\ et al.\ 2015,
   Binary Candidates in the Jovian Trojan and Hilda Populations from NEOWISE Light Curves,
   ApJ 799, 191, 20 pp
\bibitem[2012]{szabo12}
   Szab\'o, Gy.\ M., Kiss, L.\ L., P\'al, A.\ et al.\ 2012,
   Evidence for Fresh Frost Layer on the Bare Nucleus of Comet Hale-Bopp at 32 AU Distance,
   ApJ 761, 8, 7 pp
\bibitem[2016]{szabo16}
   Szab\'o, R.\ M., P\'al, A., S\'arneczky, K.\ et al.\ 2016,
   Uninterrupted optical light curves of main-belt asteroids from the K2 Mission,
   A\&A 596, A40, 9 pp
\bibitem[2017]{szabo17}
   Szab\'o, Gy.\ M., P\'al, A., Kiss, Cs.\ et al.\ 2017,
   The heart of the swarm: K2 photometry and rotational characteristics of 56 Jovian Trojan asteroids,
   A\&A 599, A44, 13 pp
\bibitem[2012]{vilenius12}
   Vilenius, E., Kiss, C., M\"uller, T.\ G.\ et al.\ 2012,
   "TNOs are Cool": A survey of the trans-Neptunian region
   VI. Herschel/PACS observations and thermal modeling of 19
   classical Kuiper belt objects
   A\&A 541, A94, 17 pp
\bibitem[2015]{viikinkoski15}
   Viikinkoski, M., Kaasalainen, M., \v{D}urech, J.\ 2015, 
   ADAM: a general method for using various data types in asteroid reconstruction,
   A\&A 576, A8, 11 pp
\bibitem[2003]{vokrouhlicky03}
   Vokrouhlicky, D., Nesvorny, D., Bottke, W.\ F.\ 2003, 
   The vector alignments of asteroid spins by thermal torques,
   Nature, 425, 147--151
\bibitem[2009]{warner09}
   Warner, B.\ D., Harris, A.\ W., Pravec, P.\  2009,
   The asteroid lightcurve database,
   Icarus 202, 134--146
\bibitem[2015]{waszczak15}
   Waszczak, A., Chang, C.-K., Ofek, E. O.\ et al.\ 2015,
   Asteroid Light Curves from the Palomar Transient Factory Survey:
   Rotation Periods and Phase Functions from Sparse Photometry,
   AJ 150, id.\ 75, 35 pp
\bibitem[2013]{zacharias13}
   Zacharias, N., Finch, C.\ T., Girard, T.\ M.\ et al.\ 2013,
   The Fourth US Naval Observatory CCD Astrograph Catalog (UCAC4),
   AJ 145, id.\ 44, 14 pp
\bibitem[2015]{zacharias15}
   Zacharias, N., Finch, C.\ T., Subasavage, J.\ P.\ et al.\ 2015,
   The First U.S. Naval Observatory Robotic Astrometric Telescope Catalog,
   AJ 150, id.\ 101, 13 pp
\end{thebibliography}

\end{document}